\documentclass{emulateapj}
\usepackage{natbib}
\newcommand{\tx}[1]{\textrm{#1}}

\newcommand{\kms}{km~$\tx{s}^{-1}$}

\def\xc{\mathrel{x_{\rm c}}}
\def\yc{\mathrel{y_{\rm c}}}
\def\rc{\mathrel{r_{\rm core}}}
\def\rt{\mathrel{r_{\rm cut}}}
\def\sigo{\mathrel{\sigma_{\rm o}}}

\slugcomment{ApJ, accepted}
\shorttitle{Lensing and Dynamical Analysis of Abell 383 and MS2137-23}
\shortauthors{Sand et al.}

\begin{document}
\title{Separating Baryons and Dark Matter in Cluster Cores: A Full 2-D Lensing and Dynamic Analysis of Abell 383 and MS2137-23}

\author{David J. Sand,$\!$\altaffilmark{1,2} Tommaso Treu\altaffilmark{3,4}, Richard S. Ellis\altaffilmark{5}, Graham P. Smith\altaffilmark{6} \& Jean-Paul Kneib\altaffilmark{7}}

\begin{abstract}
We utilize existing imaging and spectroscopic data for the galaxy
clusters MS2137-23 and Abell 383 to present improved measures of the
distribution of dark and baryonic material in the clusters' central
regions.  Our method, based on the combination of gravitational
lensing and dynamical data, is uniquely capable of separating the
distribution of dark and baryonic components at scales below 100
kpc. Our mass models include pseudo-elliptical generalized NFW
profiles for constraining the inner dark matter slope, and our lens
modeling takes into account both the ellipticity and substructure
associated with cluster galaxies as necessary in order to account for
the detailed properties of multiply-imaged sources revealed in Hubble
Space Telescope images.  We find a variety of strong lensing models
fit the available data, including some with dark matter profiles as
steep as expected from recent simulations. However, when combined with
stellar velocity dispersion data for the brightest member, shallower
inner slopes than predicted by numerical simulations are preferred, in
general agreement with our earlier work in these clusters.  For
Abell 383, the preferred shallow inner slopes are statistically a good
fit only when the multiple image position uncertainties associated
with our lens model are assumed to be 0\farcs5, to account for unknown
substructure.  No statistically satisfactory fit was obtained matching
both the multiple image lensing data and the velocity dispersion
profile of the brightest cluster galaxy in MS2137-23.  This suggests
that the mass model we are using, which comprises a pseudo-elliptical
generalized NFW profile and a brightest cluster galaxy component may
inadequately represent the inner cluster regions.  This may plausibly
arise due to halo triaxiality or by the gravitational interaction of
baryons and dark matter in cluster cores.  The intriguing results for
Abell 383 and MS2137-23 emphasize the need for a larger sample of
clusters with radial arcs. However, the progress made via this
detailed study highlights the key role that complementary observations
of lensed features and stellar dynamics offer in understanding the
interaction between dark and baryonic matter on non-linear scales in
the central regions of clusters.

\end{abstract}
\altaffiltext{1}{Chandra Fellow}
\altaffiltext{2}{Steward Observatory,
University of Arizona, 933 North Cherry Avenue, Tucson, AZ 85721}
\altaffiltext{3}{University of California, Santa Barbara, Department of Physics, Santa Barbara CA 93106-9530}
\altaffiltext{4}{Alfred P. Sloan Research Fellow}
\altaffiltext{5}{California Institute of Technology, Department of
      Astronomy, Mail Code 105--24, Pasadena, CA 91125, USA} 
\altaffiltext{6}{School of Physics and Astronomy, University of Birmingham, Edgbason, Birmingham, B15 2TT, UK}
\altaffiltext{7}{OAMP, Laboratoire d\'Astrophysique de Marseille - UMR 6110 - Traverse du siphon, 13012 Marseille, France}
\setcounter{footnote}{6}

\section{Introduction}

Cold dark matter (CDM) simulations (both with and without the
inclusion of baryonic physics) are a crucial tool and proving ground
for understanding the physics of the universe on nonlinear scales.
One of the most active aspects of research in this area concerns the
form of the dark matter density profile. Key questions raised in
recent years include: Is there a universal dark matter density profile that 
spans a wide range of halo masses? What is the form of this profile
and how uniform is it from one halo to another? To what extent
do baryons modify the dark matter distribution?

Drawing on a suite of N-body simulations, \citet{NFW97} originally proposed
that the dark matter density profiles in halos ranging in size from those
hosting dwarf galaxies to those with galaxy clusters have a
universal form. This 3-D density distribution, termed the NFW profile, 
follows $\rho_{DM}\propto r^{-1}$ within some scale radius, $r_{sc}$, and 
falls off as $\rho_{DM}\propto r^{-3}$ beyond.  Subsequent simulations 
indicated that the inner density profile could be yet steeper - 
$\rho_{DM}\propto r^{-1.5}$ \citep{M98,Ghigna00}.  As computing power 
increases and numerical techniques improve, it is now unclear whether 
the inner dark matter distribution converges to a power law form rather 
than becoming progressively shallower in slope at smaller radii
\citep{P03,Navarro04,Diemand04,Diemand05}.  

For comparisons with data, such simulations need to account for the 
presence of baryons. This is particularly the case in the cores of rich 
clusters. Although baryons represent only a small fraction of the overall 
cluster mass, they may be crucially important on scales comparable to 
the extent of typical brightest cluster galaxies. Much work is being done 
to understand the likely interactions between baryons and DM
\citep{Gnedin04,Nagai05,Faltenbacher05}.  These simulations will
provide refined predictions of the relative distributions of baryons
and DM. 

This paper is a further step in a series which aims to present an 
observational analog to progress described above in the numerical 
simulations. At each stage it is desirable to confront numerical 
predictions with observations. Whereas some workers have
made good progress in constraining the {\em total} density profile 
(e.g.~\citet{Broadhurst05b}), in order to address the relevance of the
numerical simulations we consider it important to develop and test 
techniques capable of  separating the distributions of dark and 
baryonic components (e.g. ~\citet{Sand02,Zappacosta06,Biviano06,Mahdavi07}).  

This paper presents a refined version of the method first proposed by
\citet{Sand02}, exploited more fully in \citet{Sand04} (hereafter
S04). S04 sought to combine constraints from the velocity dispersion
profile of a central brightest cluster galaxy (BCG) with a strong
gravitational lensing analysis in six carefully selected galaxy
clusters in order to separate the baryonic and dark matter
distributions.  S04 carefully selected clusters to have simple,
apparently 'relaxed' gravitational potentials in order to match
broadly the 'equilibrium' status of the simulated dark matter halos
originally analyzed by \citet{NFW97} and subsequent simulators.  For
example, Abell 383 and MS2137-23 have almost circular BCGs ($b/a$=0.90
and 0.83 respectively), require a single cluster dark matter halo to
fit the strong lensing constraints (in contrast to the more typical
clusters that require a multi-modal dark matter morphology -- Smith et
al. 2005), have previously published lens models with a relatively
round dark matter halo ($b/a$=0.88 and 0.78 respectively - Smith et
al. 2001; Gavazzi 2005), and display no evidence for dynamical
disturbance in the X-ray morphology of the clusters (Smith et
al. 2005; Schmidt \& Allen 2006). 

The merit of the approach resides in combining two techniques that
collectively probe scales from the inner $\sim$10 kpc (using the BCG
kinematics) to the $\sim$100 kpc scales typical of strong
lensing. Whereas three of the clusters contained tangential arcs,
constraining the total enclosed mass within the Einstein radius, three
contained both radial and tangential gravitational arcs, the former
providing additional constraints on the derivative of the total mass
profile. In their analysis, S04 found the gradient of the inner dark
matter density distribution varied considerably from cluster to
cluster, with a mean value substantially flatter than that predicted
in the numerical simulations.

S04 adopted a number of assumptions in their analysis whose effect on
the derived mass profiles were discussed at the time. The most
important of these included ignoring cluster substructure and adopting
spherically-symmetric mass distributions centered on the BCG. The
simplifying assumptions were considered sources of systematic
uncertainties, of order 0.2 on the inner slope. Although the six
clusters studied by S04 were carefully chosen to be smooth and round,
several workers attributed the discrepancy between the final results
and those of the simulations as likely to arise from these simplifying
assumptions \citep{Bartelmann04,Dalal04b,Meneghetti05}.

The goal of this paper is to refine the data analysis for two of the
clusters (MS2137-23 and Abell 383) originally introduced by S04 using
fully 2-D strong gravitational lensing models, thus avoiding any
assumptions about substructure or spherical symmetry. The lensing
models are based on an improved version of the LENSTOOL program
(\citealt{Kneibphd,Kneib96}; see Appendix;
http://www.oamp.fr/cosmology/lenstool/). A major development is the
implementation, in the code, of a pseudo-elliptical parameterization
for the NFW mass profile, i.e. a generalization of those seen in CDM
simulations, viz:
\begin{equation}\label{eqn:gnfw}
\label{eq:gnfw}
\rho_d(r)=\frac{\rho_{c} \delta_{c}}{(r/r_{sc})^{\beta}(1+(r/r_{sc}))^{3-\beta}}
\end{equation}
where the asymptotic DM inner slope is $\beta$. This formalism allows 
us to overcome an important limitation of previous work and takes into 
account the ellipticity of the DM halo and the presence of galaxy-scale 
subhalos. Furthermore the 2-D lensing model fully exploits the 
numerous multiply-imaged lensing constraints available for MS2137-23 
and Abell 383.

The combination of gravitational lensing and stellar dynamics is the
most powerful way to separate baryons and dark matter in the inner
regions of clusters.  However, it is important to keep a few caveats
in mind.  Galaxy clusters are structurally heterogeneous objects that
are possibly not well-represented by simple parameterized mass
models. To gain a full picture of their mass distribution and the
relative contribution of their major mass components will ultimately
require a variety of measurements applied simultaneously across a
range of radii.  Steps in this direction are already being made with
the combined use of strong and weak gravitational lensing
(e.g.~\citet{Limousin06,Bradac06}), which may be able to benefit
further from information provided from X-ray analyses
(e.g.~\citet{Schmidt06}) and kinematic studies (e.g.~\citet{Lokas03}).
A recent analysis has synthesized weak-lensing, X-ray and
Sunyaev-Zeldovich observations in the cluster Abell 478 -- similar
cross-disciplinary work will lend further insights into the mass
distribution of clusters \citep{Mahdavi07}.

Of equal importance are mass models with an appropriate amount of 
flexibility and sophistication.  For instance, incorporating models that 
take into account the interaction of baryons and dark matter may shed 
light into the halo formation process and provide more accurate 
representations of dark matter structure.  Halo triaxiality, multiple 
structures along the line of sight and other geometric effects will 
also be important to characterize. At the moment, incorporating these 
complexities and securing good parameter estimates is computationally 
expensive even with sophisticated techniques such as the Markov-Chain 
Monte Carlo method. 

Numerical simulation results are often presented as the average
profile found in the suite of calculations performed. Instead, the
distribution of inner slopes would be a more useful quantity for
comparison with individual cluster observations.  Also, comparisons
between simulations and observations would be simplified if {\it
projected} density profiles of simulated halos along multiple lines of
sight were to be made available.  These issues should be resolvable as
large samples of observed mass profiles are obtained.

For the reasons above, comparing observational results with
numerical simulations is nontrivial.  The observational task should
be regarded as one of developing mass modeling techniques of
increasing sophistication that separate dark and baryonic matter, 
so as to provide the most stringent constraints to high resolution
simulations which include baryons as they also increase in
sophistication.  The combination of stellar dynamics and strong
lensing is the first crucial step in this process.  Its diagnostic
power will be further enhanced by including other major mass
components (i.e.~the hot gas of the intracluster medium or the stellar
contribution from galaxies) out to larger radii.

A plan of the paper follows.  In \S~\ref{sec:methods} we explain the
methodology used to model the cluster density profile by combining
strong lensing with the BCG velocity dispersion profile.  In
\S~\ref{sec:obsresults} we focus on translating observational
measurements into strong lensing input parameters.  This section
includes the final strong lensing interpretation of MS2137-23 and
Abell 383.  In \S~\ref{sec:stronglens} we present the results of our
combined lensing and dynamical analysis.  In \S~\ref{sec:systematics}
we discuss further systematic effects, limitations and degeneracies
that our technique is susceptible to -- with an eye towards future
refinements.  Finally, in \S~\ref{sec:finale} we summarize and discuss
our conclusions. Throughout this paper, we adopt $r$ as the radial
coordinate in 3-D space and $R$ as the radial coordinate in 2-D
projected space.  When necessary, we assume $H_{0}$=65 \kms
Mpc$^{-1}$, $\Omega_{m}$=0.3, and $\Omega_{\Lambda}$=0.7.

\section{Methods}\label{sec:methods}

The intent of this work is to use the full 2D information provided by
the deep Hubble Space Telescope (HST) imaging in two strong lensing
clusters (MS2137-23 and Abell 383) in conjunction with the BCG stellar
velocity dispersion profile in order to constrain the distribution of
baryonic and dark matter.  These two clusters were selected for
further study from the larger sample presented by S04 because, of the
three systems with both radial and tangential gravitational arcs,
these two presented the shallowest DM inner slopes.

\subsection{Lens Modeling}\label{sec:massmodel}

We use the updated LENSTOOL ray-tracing code to construct models of
the cluster mass distribution.  Our implementation of the mass
profiles is identical to that of \citet{Golse02}, with the exception
that we have generalized their pseudo-elliptical parameterization to
include ones with arbitrary inner logarithmic slopes.  For the
details, the reader is referred to both \citet{Golse02}, and the
Appendix. Here we briefly explain the lens modeling process and
parameterization of the cluster mass model.

Identifying mass model components and multiple-image candidates is an
iterative process. Initially, multiple images are spectroscopically
confirmed systems with counter images identified by visual inspection
and with the aid of preliminary lens models, taking into account that
gravitational lensing conserves surface brightness.  Multiple images
without spectroscopic confirmation were used in the case of Abell 383,
since these additional constraints helped clarify the role that galaxy
perturber \#1 (Table~\ref{tab:lensmodel}) played in the central regions
of the cluster (see \S~\ref{sec:lensinterpa383}).  If the location of
a counter image is tentative, especially if there are several
possibilities or an intervening cluster galaxy confuses the situation,
the system is not included in deriving the mass model.
\S~\ref{sec:lensinterpms2137} and \S~\ref{sec:lensinterpa383} present
a detailed description of the final multiple image list adopted.

Once the multiple images are determined, the cluster mass model is
refined and perturber galaxy properties are fixed.  In general, a lens
mass model will have both cluster and galaxy scale mass components.
The cluster scale mass component represents the DM associated with the
cluster as a whole plus the hot gas in the intracluster medium.  In
the limit that the cluster DM halo is spherical (see
Eqn~\ref{rho_ell}), its density profile has the form of
Eqn~\ref{eqn:gnfw}.  In the adopted parameterization, the DM halo also
has a position angle ($\theta$) and associated pseudo-ellipticity
($\epsilon$) (see Eqn~\ref{sigma_ell} \& \ref{rho_ell}).

Galaxy scale mass components are necessary to account for
perturbations to the cluster potential that seem plausible based on
the HST imaging and are demanded by the observed multiple image
positions.  These components are described by pseudo-isothermal
elliptical mass distributions (PIEMD; \citet{Kassiola93}).  Each PIEMD
mass component is parametrized by its position ($\xc$, $\yc$),
ellipticity ($e$), position angle ($\theta$), core radius ($\rc$),
cut-off radius ($\rt$) and central velocity dispersion ($\sigo$).  The
projected mass density, $\Sigma$, is given by:

\begin{equation}\label{eq:piemd}
\Sigma(x,y){=}\frac{\sigo^2}{2G}\,\frac{\rt}{\rt{-}\rc}\left(\frac{1}{(\rc^2{+}\rho^2)^{1/2}}{-}\frac{1}{(\rt^2{+}\rho^2)^{1/2}}\right)
\end{equation}

\noindent where
$\rho^2{=}[(x{-}\xc)/(1{+}e)]^2{+}[(y{-}\yc)/(1{-}e)]^2$
and the ellipticity of the lens is defined as $e{=}(a{-}b)/(a{+}b)$
\footnote{This quantity should not be confused with the quite
different definition used for the pseudo-elliptical generalized NFW
profile, see the Appendix.}.  The total mass of the PIEMD is thus
$3/2\pi \sigma_0^2 r_{\rm cut}/G$. In order to relate
Equation~\ref{eq:piemd} to the observed surface brightness of the BCG
in particular, we take $\Sigma=(M_*/L)I$, where $M_*/L$ is the stellar
mass to light ratio and $I$ is the surface brightness, and find the
following relation

\begin{equation} \label{eq:mlpiemd}
M_*/L = 1.50 \pi\sigo^{2}\rt / (GL)
\end{equation}

\noindent where $L$ is the total luminosity of the BCG.  The $M_*/L$
of the central BCG will be used as a free parameter in our mass
modeling analysis.  Further details and properties of the truncated
PIEMD model can be found in \citet{Natarajan97} and
\citet{Limousin05}.

The relevant parameters of the perturber galaxies (position,
ellipticity, core radius, cutoff radius and position angle) are
assumed to be those provided via examination of the HST imaging (see
\S~\ref{sec:galgeom} for details). Only the central stellar velocity
dispersion, $\sigo$, is determined by optimization. At a particular
stage in the process, the predicted multiple image positions are
compared with those observed, and a $\chi^{2}$ value calculated (see
\S~\ref{sec:stats}).  The iteration stops when a $\chi^{2}$ minimum is
reached.  The sole criteria for adding a perturber galaxy was
whether or not it was necessary for the lens model to match the
multiple image positions.  If adding an additional perturber did not
alter our interim minimum $\chi^{2}$, we did not include it in our
subsequent analysis.

Two special cases were encountered during the above procedure.
First, one of the galaxy perturbers in Abell~383 required a larger
cutoff radius ($r_{cut}$) than implied from the light distribution.
As is described in \S~\ref{sec:lensinterpa383}, this concentration is
necessary to account for several of the multiple image positions.  For
this mass concentration, not only do we determine the optimum
$\sigma_{0}$ parameter, but also the cutoff radius ($r_{cut}$).

The other special case concerns the BCG in both galaxy clusters.
These are assumed to be coincident with the center of the cluster DM
halo -- one justified by the co-location of the BCG and central X-ray
isophotes \citep{gps05,gavazzi03}.  The BCG mass distribution,
represented by a PIEMD model, comprises only the stellar mass.  In
this case the HST imaging is used to fix the BCG ellipticity and
position angle, but since the measured stellar velocity dispersion is
to be used as a constraint on the cluster mass profile, we leave the
central velocity dispersion parameter (and hence the stellar $M_*/L$,
Eqn~\ref{eq:mlpiemd}) free {\it in the lensing analysis}.  As the
Jaffe density profile is used for the BCG dynamical analysis (S04),
the PIEMD core and cutoff radius which best match the Jaffe surface
density are adopted.  \S~\ref{sec:galgeom} discusses further the
results of the surface brightness matching between the Jaffe and PIEMD
models in the clusters.

\subsection{Incorporating the Dynamical Constraints}\label{sec:dynconsts}

Apart from the use of the pseudo-elliptical gNFW profile for the dark
matter component, the observational data and analysis methods adopted
here are identical to those used by S04. In that work, the observed
velocity dispersion profile of the BCG was interpreted via the
spherical Jeans equation (see Appendix of S04) to assist in the
decomposition of the dark and baryonic mass components.  This portion
of the $\chi^{2}$ was calculated by comparing the expected velocity
dispersion profile of the BCG (which depends on the mass enclosed at a
given radius and the relative contribution of dark and luminous
matter) given a mass model with the observed velocity dispersion
profile, taking into account the effects of seeing and the long slit
shape used for the observations. Ellipticity in the BCG and its dark
matter halo can be ignored as its effect on the velocity dispersion
profile will be small (e.g.~\citet{Kronawitter00}).

The reader is referred to S04 for the observational details pertaining
to the velocity dispersion profile, the surface brightness profile of
the BCG and the subsequent dynamical modeling to constrain the cluster
DM inner slope.

\subsection{Statistical Methods}\label{sec:stats}

A $\chi^{2}$-estimator is used to constrain the acceptable range of
parameters compatible with the observational data. First we use the
strong lens model to calculate the likelihood of the lensing
constraints and then we combine it with a dynamical model to include
the kinematic information into the likelihood.

The first step is the strong lensing likelihood.  Once the lensing
interpretation is finalized and the perturbing galaxy parameters are
fixed, the remaining free parameters are constrained by calculating a
lensing $\chi^2$ over a hypercube which encompasses the full range of
acceptable models, modulo a prior placed on the dark matter scale
radius (see \S~\ref{sec:rscprior}).  In Bayesian terms this
corresponds to adopting a uniform prior.  The lensing $\chi^2$ value
is calculated in the source plane identically to that of
\citep{gps05}.  For each multiply-imaged system, the source
location for each noted image ($x_{model,i}$,$y_{model,i}$) is
calculated using the lens equation.  Since there should be only one
source for each multiple image set (with N images), the difference
between the source positions should be minimized, hence:

\begin{equation}
\chi^2_{pos}{=}\sum_{i{=}1}^{N}\frac{(x_{model,i}{-}x_{model,i+1})^2{+}(y_{model,i}{-}y_{model,i+1})^2}{\sigma_{S}^2}
\label{theory:chipos}
\end{equation}

\noindent $\sigma_{S}$ is calculated by scaling the positional error
associated with a muliple image knot, $\sigma_{I}$, by the
amplification of the source $A$ so that $\sigma_{I}^{2}=A
\sigma_{S}^{2}$.  The following analysis will assume two different
positional errors for the multiply imaged knots, using uncertainties
of $\sigma_{I}=$0\farcs2 and 0\farcs5, referred to hereafter as the
`fine' and `coarse' analyses, respectively. The case for each is
justified below.

%


The finer 0\farcs2 error bar corresponds to the uncertainty in the
multiple image knot positions as defined by the resolution and pixel
size of the HST/WFPC2 images.  Excellent strong lensing fits are
achieved with the finer 0\farcs2 error bar ($\chi^{2}/dof\sim 1$), so
that technically there is no need for increased uncertainties.  The
uncertainty is dominated by the spatial extension of the multiple
image knots employed and the ability to identify surface brightness
peaks.

In contrast to our ability to match the image positions down to the
resolution of the HST/WFPC2 images, recent combined strong and weak
lensing analyses of Abell 1689 have been unable to do so
(e.g.~\citet{Broadhurst05,Halkola06,Limousin06}).  Although Abell 1689
is a more complex cluster than those studied here, it does have the
most identified multiple images of any other cluster to date, and so
can probe the overall mass profile on smaller scales and with many
more constraints.  As espoused in \S~1, real galaxy clusters are
complex systems that are likely not easily parameterized by simple
mass models, and as the number and density of mass probes increases
the more refined and complete the mass model necessary to match the
data.  In our case, where we have relatively few mass profile
constraints (at least in comparison to Abell 1689), we adopt a coarser
0\farcs5 positional error which allows us to account for complexities
in the actual mass distribution of our clusters that our small number
of mass probes are insensitive to.  This, plus the fact that we
carefully chose our perturbing galaxies such that a lensing
$\chi^{2}/dof\sim1$ was found should account for reasonable situations
where we have missed an interesting perturber galaxy.  By adopting
too small a multiple image position uncertainty, the region in
parameter space explored may be overly confined (such that, for
example, the observed BCG velocity dispersion profile cannot be
reproduced).

The strong lensing analysis is performed with 5 free parameters,
analogous to those adopted in S04.  These are the dark matter inner
logarithmic slope ($\beta$), the pseudo-ellipticity of the potential
($\epsilon$), the amplitude of the DM halo ($\delta_{c}$), the dark
matter scale radius ($r_{sc}$), and the mass-to-light ratio of the BCG
($M_*/L$).  We choose to place a uniform prior on the dark matter
scale radius, ($r_{sc}$) based on past mass profile analyses of these
clusters and results from CDM simulations in order to reduced
computation time (see \S~\ref{sec:rscprior}).  In practice, to
evaluate the $\chi^2_{pos}$ at each point in the hypercube, the
pseudo-ellipticity of the cluster dark matter halo is optimized while
simply looping over the remaining free parameters.

Once the strong lensing $\chi^{2}$ values are computed, attention is
turned to the dynamical data.  In contrast to the strong lensing
model, the dynamical model is spherically symmetric and follows that
presented by S04, with the $\chi^{2}$ value being:

\begin{equation}
\chi^{2}_{\sigma}{=}\sum_{i{=}1}^{N}\frac{(\sigma_{i,obs}{-}\sigma_{i,model})^{2}}{\Delta_{i}^{2}}
\label{eqn:chisigma}
\end{equation}

\noindent where $\Delta_{i}$ is the uncertainty in the observed
velocity dispersion measurements.  

The lensing and velocity dispersion $\chi^{2}$ values are summed, 
allowing for standard marginalization of nuisance parameters and 
the calculation of confidence regions.

\subsection{Dark matter scale radius prior}\label{sec:rscprior}

As mentioned in the previous sections, in order to limit computation
time, a prior was placed on the dark matter scale radius
$r_{sc}$. This is justified both on previous mass profile analyses of
these clusters and the results of CDM simulations.

An array of CDM simulations has provided information not only on the
inner dark matter density profile, but on the expected value of the
scale radius, $r_{sc}$, and its intrinsic scatter at the galaxy
cluster scale (e.g.~\citealt{Bullocketal01,Tasitsiomi04,Dolag04}).
For example, \citet{Bullocketal01} found that dark matter halos the
size of small galaxy clusters have scale radii between 240 and 550 kpc
(68\% CL).  \citet{Tasitsiomi04}, using higher resolution simulations
with fewer dark matter halos found $r_{sc}$ of 450$\pm$300 kpc.
Finally, \citet{Dolag04} studied the DM concentrations of galaxy
clusters in a $\Lambda CDM$ cosmology and found a typical range of
scale radii between $r_{sc}$ of 150 and 400 kpc.  These results
represent a selection of the extensive numerical work being done on
the concentration of dark matter halos.

Previous combined strong and weak lensing analyses of MS2137-23 have
provided approximate values for the scale radius
\citep{gavazzi03,Gavazzi05}.  \citet{gavazzi03} found a best fitting
scale radius of $\sim$130 kpc (and hints that the scale radius may be
as low as $\sim$70 kpc from their weak lensing data) for their
analysis of MS2137-23.  A more recent analysis \citep{Gavazzi05} found a
best fitting radius of $\sim$170 kpc. Similarly for Abell 383, a
recent X-ray analysis found a best-fitting dark matter scale radius of
$\sim130$ kpc (Zhang et al. 2007).

Taking these observational studies into account, we chose a uniform
scale radius prior between 100 and 200 kpc for MS2137-23 and Abell
383, both for simplicity and to bracket the extant observational
results which often have constraints at larger radii (and thus
constrain the scale radius better) than the current work.  It is worth
noting that the extant observations of these two clusters indicate a
scale radius which is on the low end of that predicted from CDM
simulations.  For a fixed virial mass, a smaller scale radius
indicates a higher concentration, $c=r_{vir}/r_{sc}$.  This could be
due to the effect of baryonic cooling, which could increase halo
concentration (as well as inner slope perhaps). It has been suggested
that those halos with the highest concentration (again for a fixed
mass) are those that are the oldest and with the least substructure,
providing more indirect evidence that we have chosen 'relaxed' galaxy
clusters to study \citep{Zentner05}. We briefly explore the
consequences of changing our assumed scale radius prior range in
\S~\ref{sec:highrad}.

\section{Application to Data}\label{sec:obsresults}

We now turn to the observational data for MS2137-23 and Abell 383
and describe our methods for analyzing these in the context
of lensing input parameters.

\subsection{BCG and Perturber Galaxy Parameters}\label{sec:galgeom}

In order to fix the position angle and ellipticity of the perturber
galaxies and BCG components, the IRAF task {\sc ellipse} is used to
estimate the surface brightness profile at typically the effective
radius.  The measured parameters are fixed in the lensing
analysis. The galaxy position, core radius ($\rc$) and cutoff radius
($\rt$) are each chosen to match those fitting the photometry
(Table~\ref{tab:lensmodel}).  For perturbing galaxies, this leaves
only the PIEMD parameter velocity dispersion ($\sigo$) which must be
adjusted to match the multiple imaging constraints, as explained in
\S~\ref{sec:massmodel}.

For the BCG, following S04, it is preferable to use the Jaffe stellar
density profile for the dynamical analysis since this function
provides an analytic solution to the spherical Jeans equation.
However, the PIEMD model implemented in {\sc lenstool} offers numerous
advantages for the lensing analysis. To use the most advantageous
model in each application, a correspondence is established between the
two by fitting with a PIEMD model the Jaffe surface brightness fit
presented by S04.  An appropriate combination of the PIEMD $\rc$ and
$\rt$ model parameters matches the Jaffe profile found by S04 with no
significant residuals (PSF smearing was also taken into account).
Table~\ref{tab:lensfixed} lists the PIEMD parameters used for our
lensing analysis, as well as the Jaffe profile parameters used by S04.

\begin{table*}
\begin{center}
\caption{Fixed Parameters in Abell 383 and MS2137-23 Lens Model\label{tab:lensfixed}}
\begin{tabular}{lcccccccc}
\tableline\tableline
Cluster&Parameters& $x_{c}$&$y_{c}$& b/a & $\theta$&$r_{core}$&$\sigma_{0}$& $r_{cut}/R_{e}$\\
&&(arcsec)&(arcsec)&&(deg)&(kpc)&($km s^{-1}$)&(kpc)\\
\tableline
MS2137&Cluster-scale DM halo&0.0&0.0&Free&5.0 & -&-&-\\
&${\rm BCG_{PIEMD}}$&0.0&0.0&0.83&17.75&5$\times10^{-6}$&Free&22.23\\
&${\rm BCG_{Jaffe}}$&0.0&0.0&-&-&-&-&24.80\\
&Galaxy Perturber&16.2&-5.46&0.66&159.9&0.05&173.0&4.8\\
Abell 383&Cluster-scale DM halo&0.0&0.0&Free&104.3&-&-&-\\
&${\rm BCG_{PIEMD}}$&0.0&0.0&0.90&107.2&3$\times10^{-6}$&Free&25.96\\
&${\rm BCG_{Jaffe}}$&0.0&0.0&-&-&-&-&46.75\\
&Perturber 1&14.92&-16.78&0.804&-20.9&0.67&230.0&26.98\\
&Perturber 2&9.15&-1.92&0.708&10.3&0.51&140.0&10.79\\
&Perturber 3&0.17&-24.26&0.67&65.2&0.24&124.8&9.10\\
&Perturber 4&-4.10&-13.46&0.645&27.7&0.17&125.7&2.19\\
\tableline
\end{tabular}
\tablecomments{The position angle, $\theta$ is measured from North
towards East.  The DM halo is parameterized with the pseudo-gNFW
profile.  All other mass components are parameterized by a PIEMD
model. Note that $R_{e}=0.76r_{jaffe}$ \citep{Jaffe83}.}
\end{center}
\end{table*}

\subsection{MS2137-23 Lens Model}\label{sec:lensinterpms2137}

The strong lensing properties of MS2137-23 have been studied
extensively by many workers
\citep{Mellier93,Miralda95,Hammer97,gavazzi03,Gavazzi05}.  The most
detailed model \citep{gavazzi03} used 26 multiply-imaged knots from
two different background sources.  The model adopted here is more
conservative and based only on those multiple images confirmed via
spectroscopy or suggested on the grounds of surface brightness or
interim lens models.  Despite having some multiple images in common
with \citet{gavazzi03}, we have retained our own nomenclature.

\begin{figure*}  
\begin{center}

\mbox{
\mbox{\epsfysize=6.2cm \epsfbox{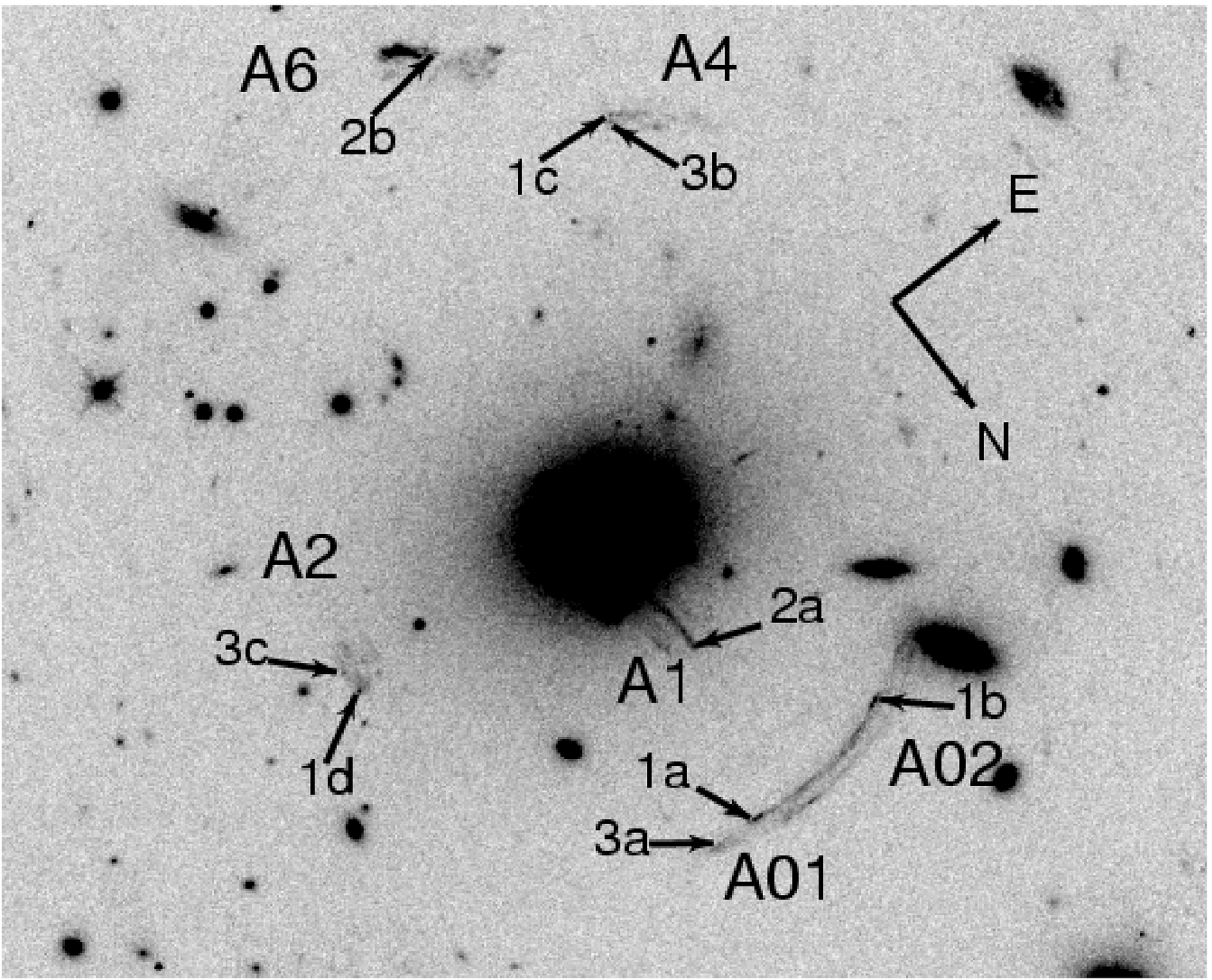}}
\mbox{\epsfysize=6.2cm \epsfbox{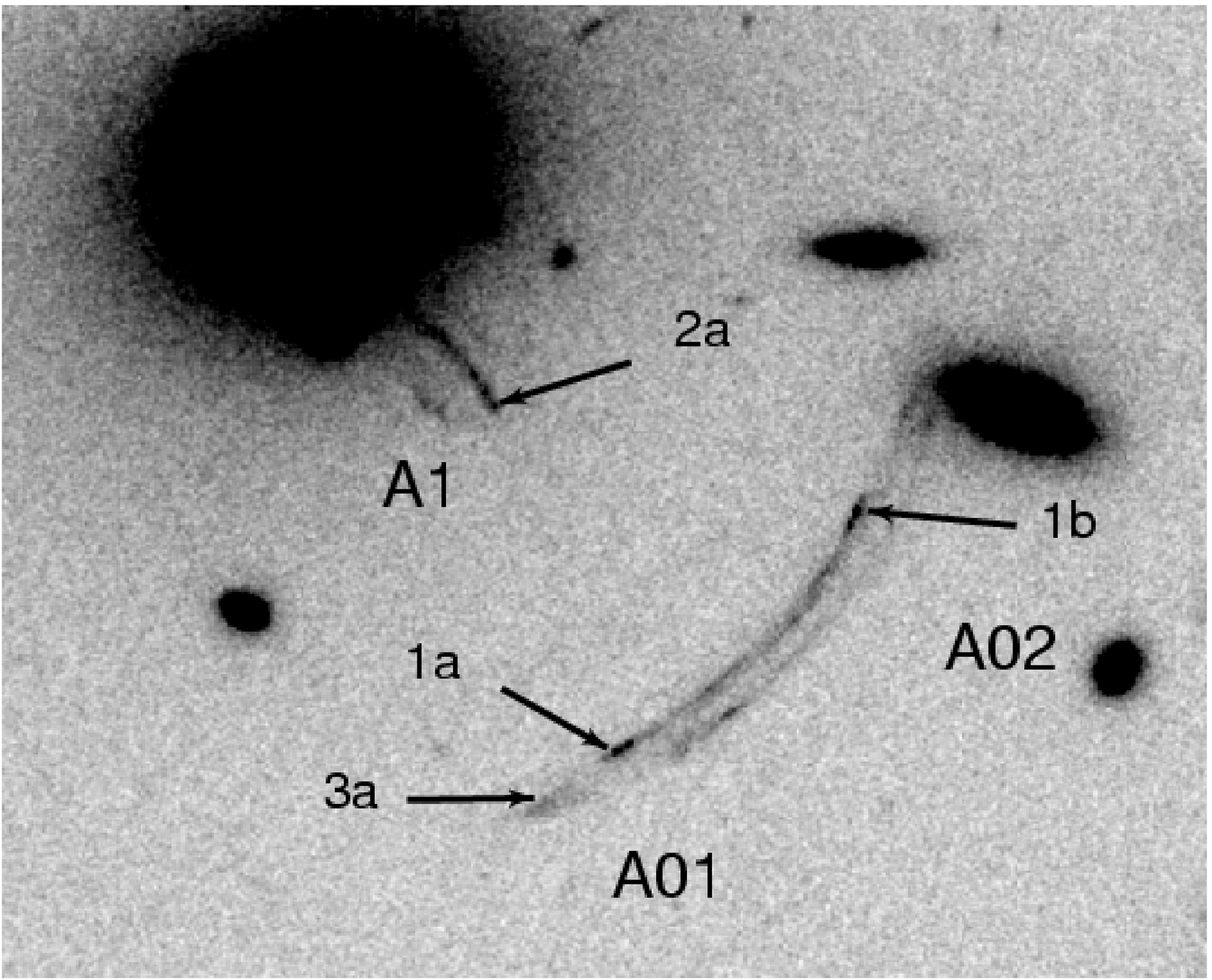}}
}
\mbox{
\mbox{\epsfysize=6.2cm \epsfbox{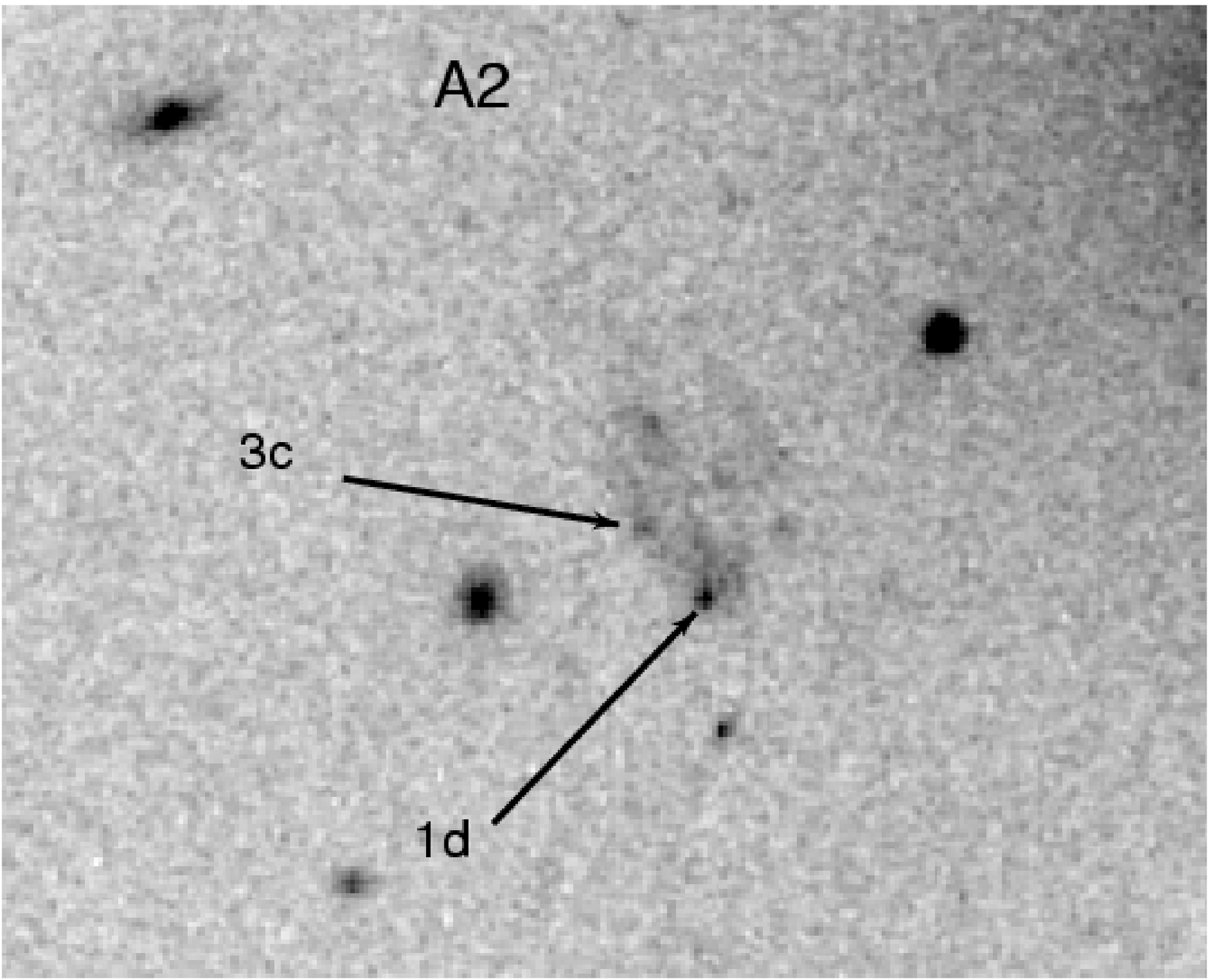}}
\mbox{\epsfysize=6.2cm \epsfbox{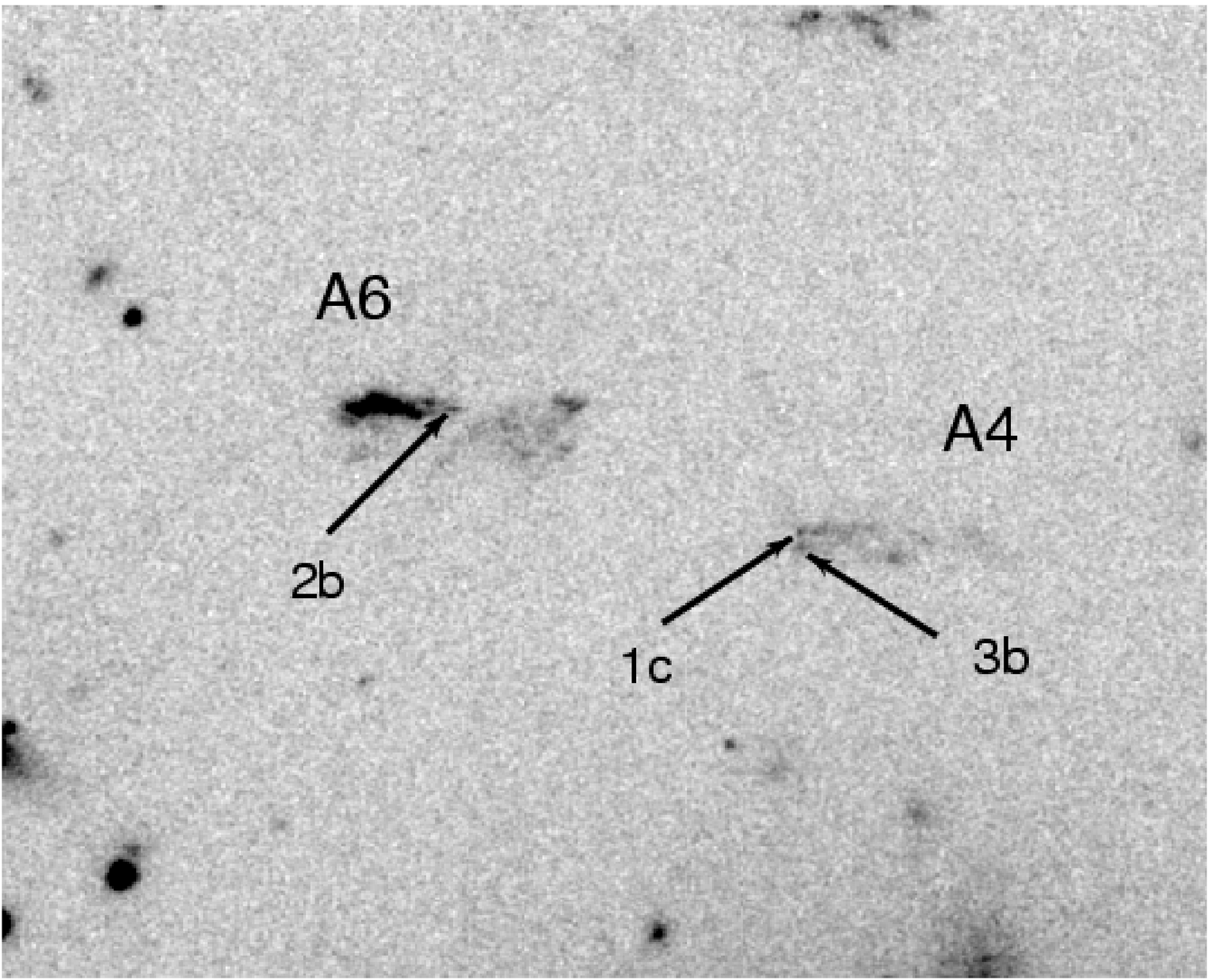}}
}

\caption[Multiple image interpretation of the cluster
MS2137.]{\label{fig:mulplot} Multiple image interpretation of the
cluster MS2137-23.  The exact positions used are shown in
Table~\ref{tab:lensinterp}.  Three sets of multiple images are
identified, one with the radial arc system (2a \& 2b) and two with the
tangential arc system (1abcd \& 3abc).  The perturbing galaxy is the
elliptical S0 next to the lensed feature 1b. }
\end{center}
\end{figure*}

Following \citet{Sand02}, the tangential and radial arcs arise from
separate sources, at $z=1.501$ and $z=1.502$, respectively.  The
multiple image interpretation is detailed in Figure~\ref{fig:mulplot}
and Table~\ref{tab:lensinterp}.  There are two separate features (1
and 3) on the source giving rise to the tangential arc which is
multiply-imaged four and three times respectively.  It has not been
possible to confidently locate the fourth image of feature 3, since it
is adjacent to the perturbing galaxy.  Also, it is expected that a
fifth, central image would be associated with the giant tangential
arc.  Although the position of this fifth central image has been
tentatively reported \citep{gavazzi03}, we do not include it in our
model because we were unable to clearly identify it due to BCG
subtraction residuals.  Two images of the source giving rise to the
radial arc were also identified.  The mirror image of feature 2a
nearer the center of the BCG could not be recovered, most likely
because of residuals arising from the subtraction of the BCG.

As in \citet{gavazzi03}, only one perturbing galaxy is included in the
lens model (see Table~\ref{tab:lensfixed}). For this system, the
best-fitting $\chi^{2}$ value occurred near $\sigo$=173 \kms.

In the initial modeling of MS2137-23, some experimentation was
undertaken with different cluster DM ellipticities and position
angles.  While some variation in ellipticity is permitted by the
lensing interpretation, a robust position angle offset was detected
between the BCG and that of the DM halo of $\Delta \theta \sim$13
degrees, in agreement with \citet{gavazzi03}.  In the following,
results are presented with the DM position angle fixed at $\theta$=5
degrees.  This optimal position angle was determined during the
initial lens modeling process by fixing all cluster mass parameters to
values corresponding to a model with $\chi^{2}/dof\sim 1$ and letting
the DM position angle vary until a $\chi^{2}$ minimum was reached.
As a consistency check, we repeated our calculations with a
fixed DM position angle of 4.0 and 6.0 degrees.  Varying the DM
position angle had very little effect on our other parameter
constraints, but results in a slightly larger overall $\chi^{2}$
(lensing + velocity dispersion profile; $\Delta \chi^{2} <$1) value.
For this reason, we only present our DM position angle of 5 degree
results.

\subsection{Abell 383 Lens Model}\label{sec:lensinterpa383}

\begin{figure*}  
\begin{center}
\mbox{
\mbox{\epsfysize=6.2cm \epsfbox{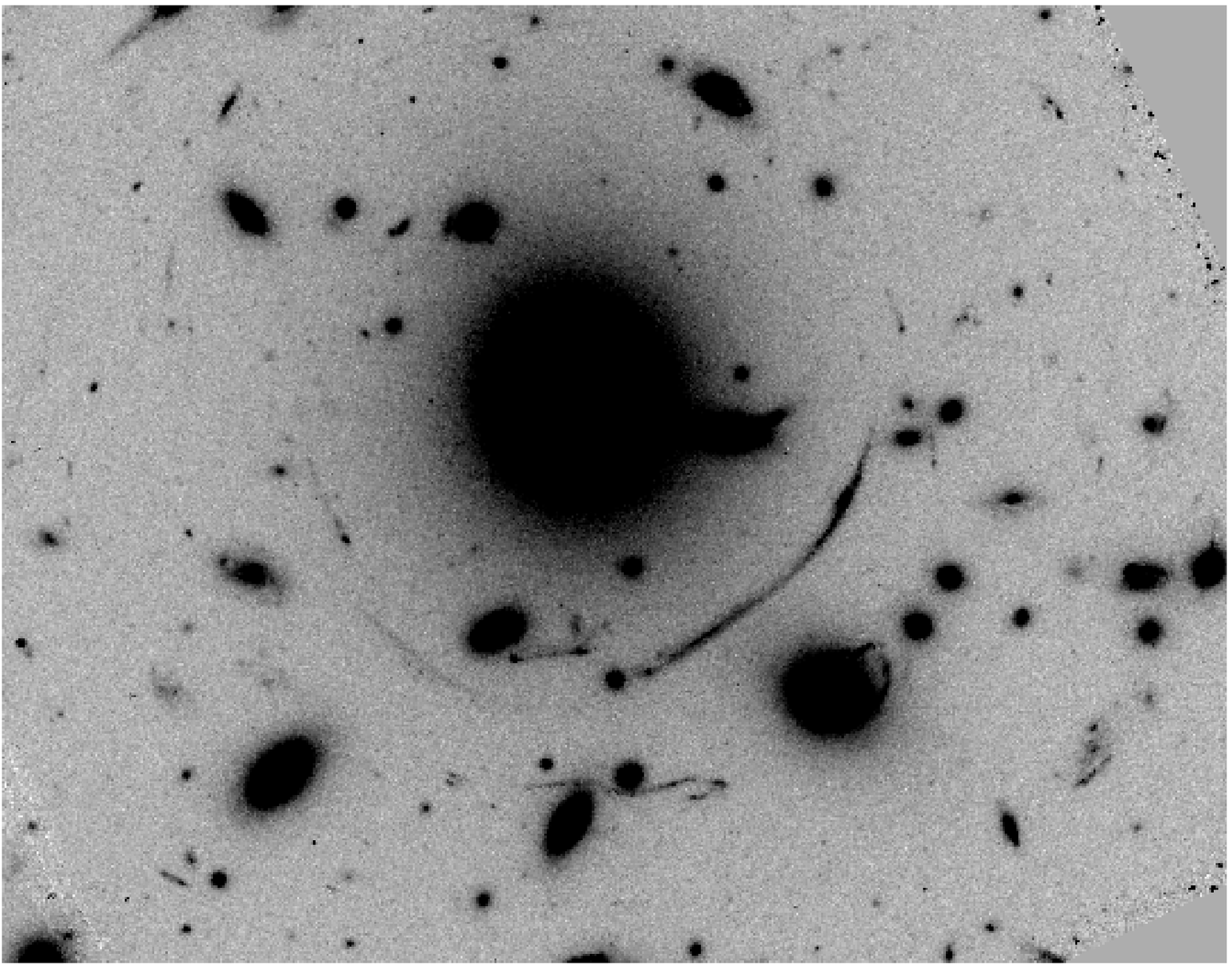}}
\mbox{\epsfysize=6.2cm \epsfbox{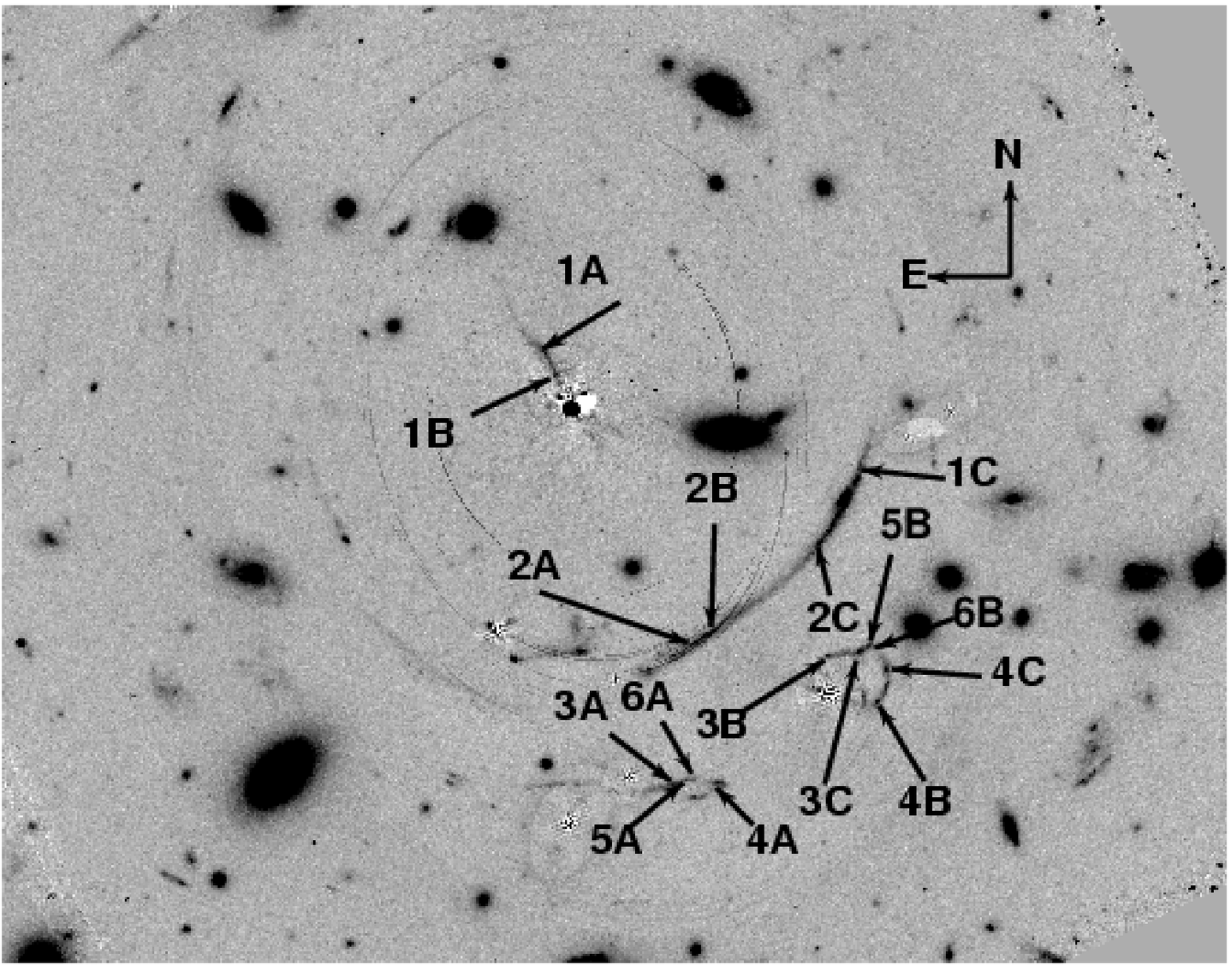}}
}
\mbox{
\mbox{\epsfysize=6.2cm \epsfbox{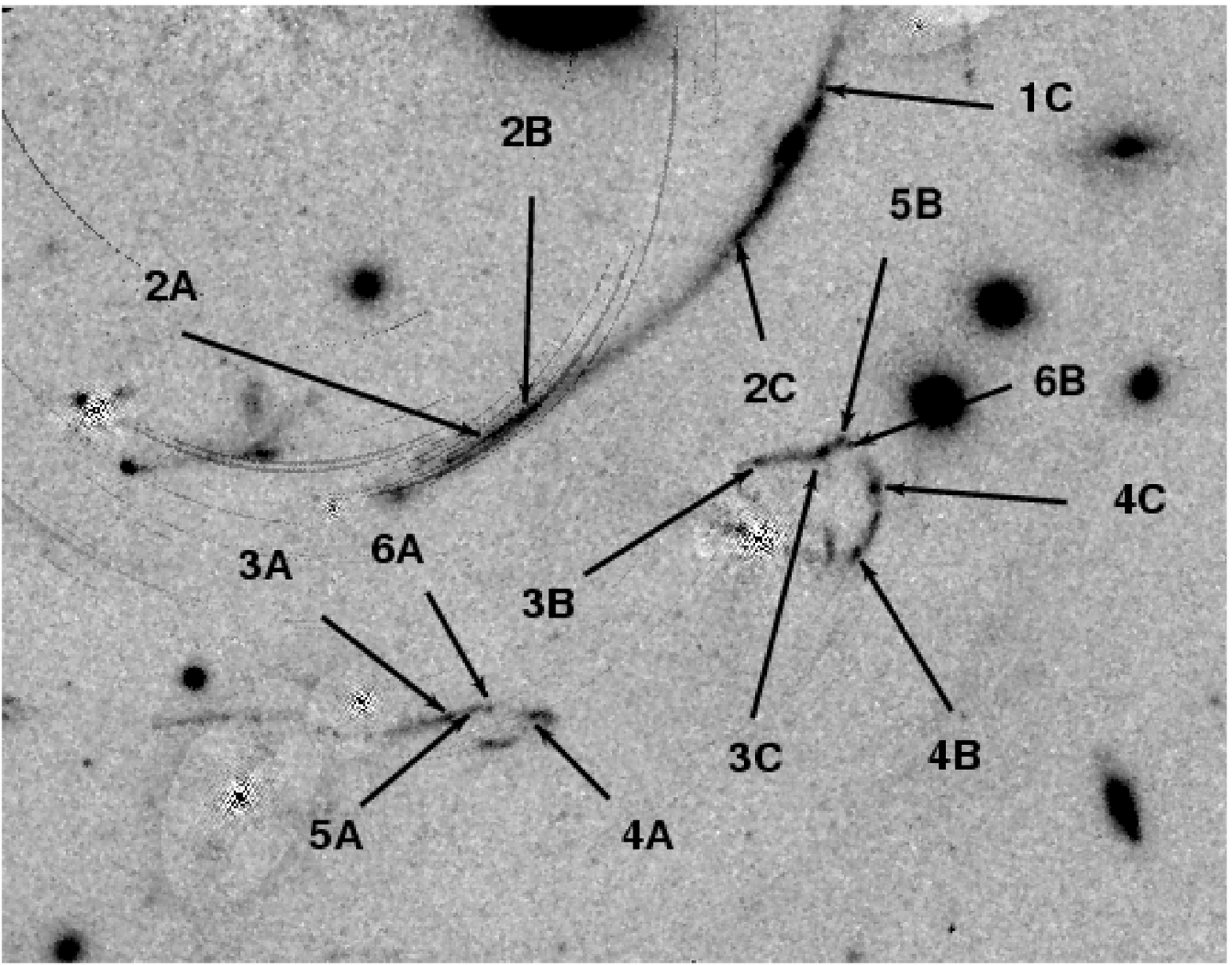}}
\mbox{\epsfysize=6.2cm \epsfbox{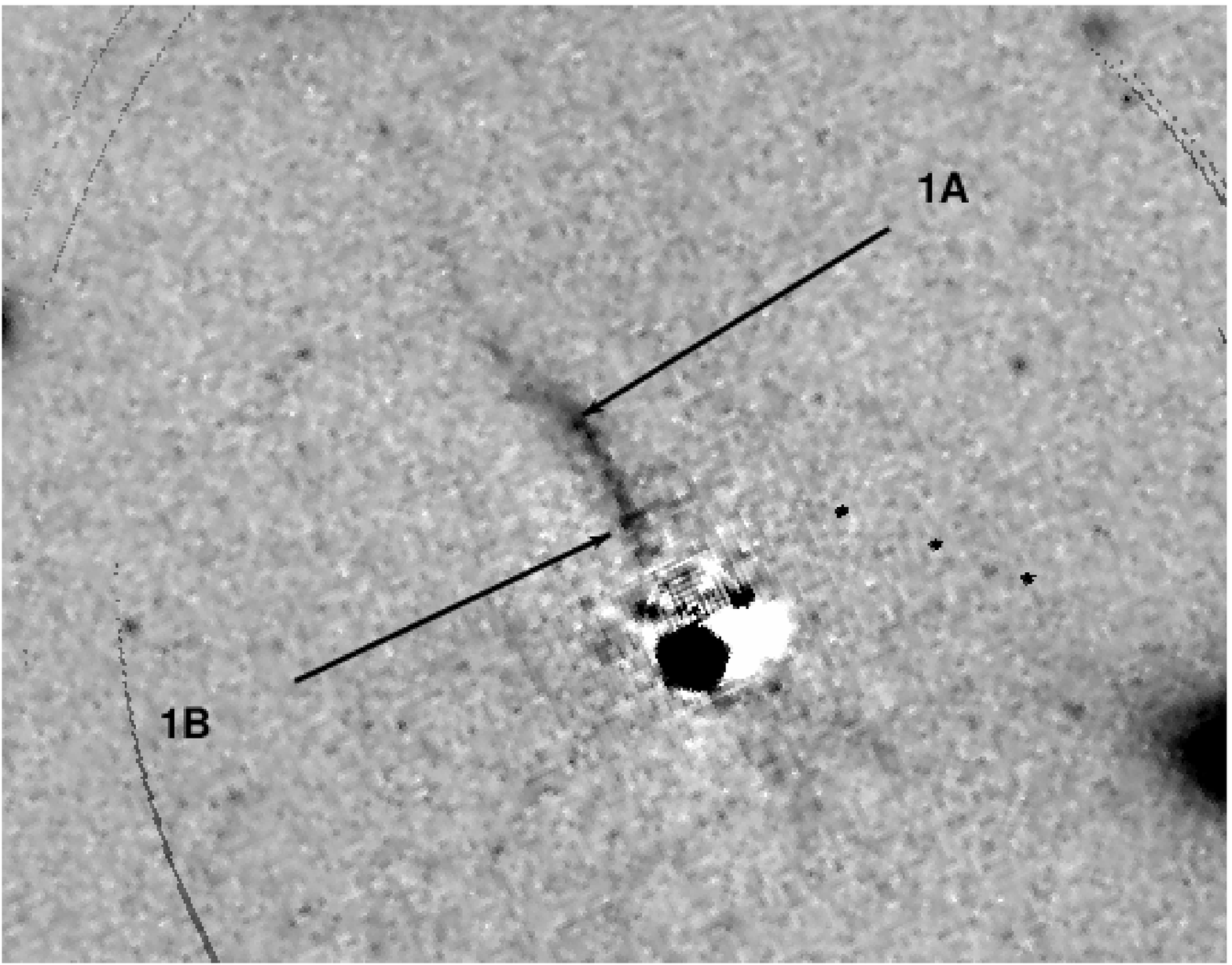}}
}
\caption[Multiple image interpretation of the cluster Abell
383.]{\label{fig:mulplota383} Multiple image interpretation of the
cluster Abell 383.  The exact positions used are shown in
Table~\ref{tab:lensinterp}.  In all figures except for the top left we
have subtracted cluster galaxies in order to more clearly see the
multiple image features. }
\end{center}
\end{figure*}

Detailed lens models for Abell 383 have been published in
\citet{gps01,gps05}, which we will largely adopt in this work.
Multiple image sets 1 and 2 are based on the in-depth lensing
interpretation of \citet{gps05}.  The reader is referred to that work
for a detailed description of this radial and tangential gravitational
arc.  Multiple image sets 3, 4, 5 and 6 (for which there are no
spectroscopic redshifts, but for which their distinctive morphology is
reassuring) are included largely to constrain the properties of
perturbing galaxies 1, 3, and 4 (see Fig~\ref{fig:mulplota383} and
Table~\ref{tab:lensinterp}).  Since these images have no spectroscopic
confirmation, a redshift $z\sim$3 was assumed; the mass model is very
insensitive to the exact choice.

The Abell 383 cluster mass model is more complex than that for
MS2137-23, but only in the sense that there are more perturbing
galaxies that must be put into the mass model to match the image
positions.  The bright cluster elliptical southwest of the BCG (\#2 in
\citet{gps05}; Perturber \#1 in this work) requires a DM halo more
extended than the light, as mentioned in \S~\ref{sec:massmodel}.
After some iteration, it was found that the parameters of this
important perturber could be fixed to those listed in
Table~\ref{tab:lensfixed}.  Multiple image sets 3, 4 and 5 play a
crucial role (see Table~\ref{tab:lensinterp}) in constraining the
perturber. Although other perturbing galaxies were added, none has a
comparable effect on the lensing $\chi^2$.  Table~\ref{tab:lensfixed}
provides the full model parameter list.

A slight ($\sim$3 degree) offset between the position angle of the BCG
and the cluster DM halo was noted.  We found the best-fitting DM
position angle in the same way as in MS2137-23
(\S~\ref{sec:lensinterpms2137}).  The position angle of the DM halo
was kept fixed but the ellipticity was left as a free parameter. As in
MS2137-23, we also reran our analysis with a DM position angle of
103.3 and 105.3, but only present the results with a DM PA of 104.3.

\begin{table*}
\begin{center}
\caption{Multiple Image Interpretation of MS2137 and Abell 383\label{tab:lensinterp}}
\begin{tabular}{lcccc}
\tableline\tableline
Cluster&Multiple Image& $x_{c}$&$y_{c}$&Redshift\\
&ID&(arcsec)&(arcsec)\\
\tableline
MS2137&1a&6.92&-13.40&1.501\\
&1b&12.40&-7.94&1.501\\
&1c&0.07&19.31&1.501\\
&1d&-11.57&-7.49&1.501\\
&2a&3.96&-5.51&1.502\\
&2b&-8.01&22.10&1.502\\
&3a&5.16&-14.68&1.501\\
&3b&0.11&18.91&1.501\\
&3c&-12.30&-6.74&1.501\\
Abell 383&1A&-1.74&2.56&1.0\\
&1B&-1.03&1.20&1.0\\
&1C&16.37&-4.03&1.0\\
&2A&7.00&-14.01&1.0\\
&2B&8.23&-13.20&1.0\\
&2C&14.11&-8.19&1.0\\
&3A&5.88&-22.02&3.0\*\\
&3B&14.69&-14.68&3.0\*\\
&3C&16.49&-14.39&3.0\*\\
&4A&8.35&-23.96&3.0\*\\
&4B&17.45&-17.28&3.0\*\\
&4C&17.92&-15.43&3.0\*\\
&5A&6.64&-21.75&3.0\*\\
&5B&16.98&-14.09&3.0\*\\
&6A&7.05&-21.75&3.0\*\\
&6B&17.27&-14.17&3.0\*\\
\tableline
\end{tabular}
\tablecomments{Multiple Image Interpretation.  All image positions are with
respect to the BCG center.}
\end{center}
\end{table*}

\section{Results}\label{sec:stronglens}

We now analyze the refined 2D lens models of MS2137-23 and Abell 383
together with the velocity dispersion profiles presented in S04.  We
present our analysis with both a multiple image position uncertainty
of 0\farcs2 and 0\farcs5, described as the fine and coarse fits in
\S~\ref{sec:stats}.  The results are summarized in
Table~\ref{tab:lensmodel} and Figures~\ref{fig:ms2137rscprior} and
\ref{fig:a383rscprior}.

\subsection{MS2137-23}

\begin{figure*}
\begin{center}
\mbox{
\mbox{\epsfysize=4.5cm \epsfbox{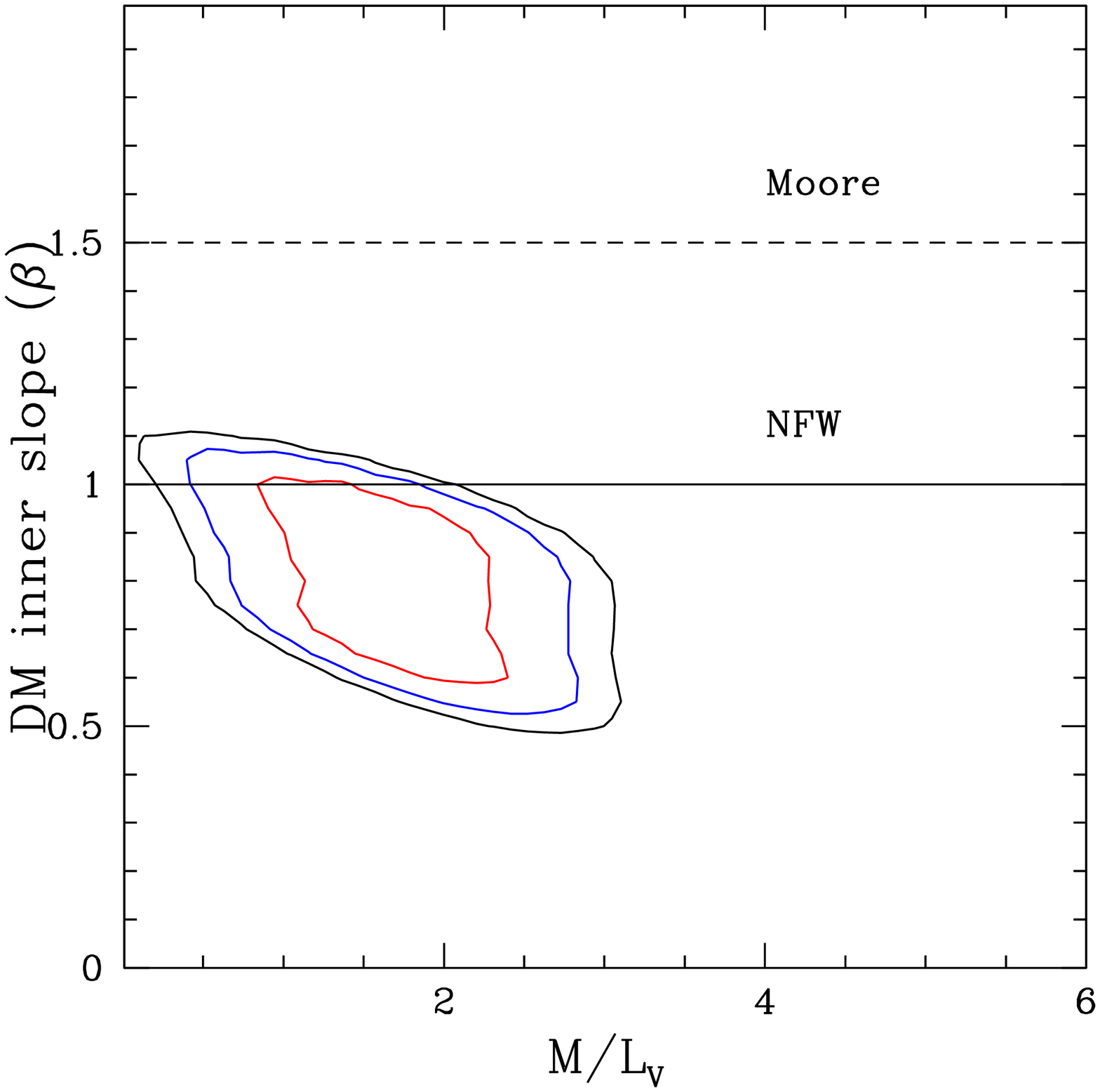}}
\mbox{\epsfysize=4.5cm \epsfbox{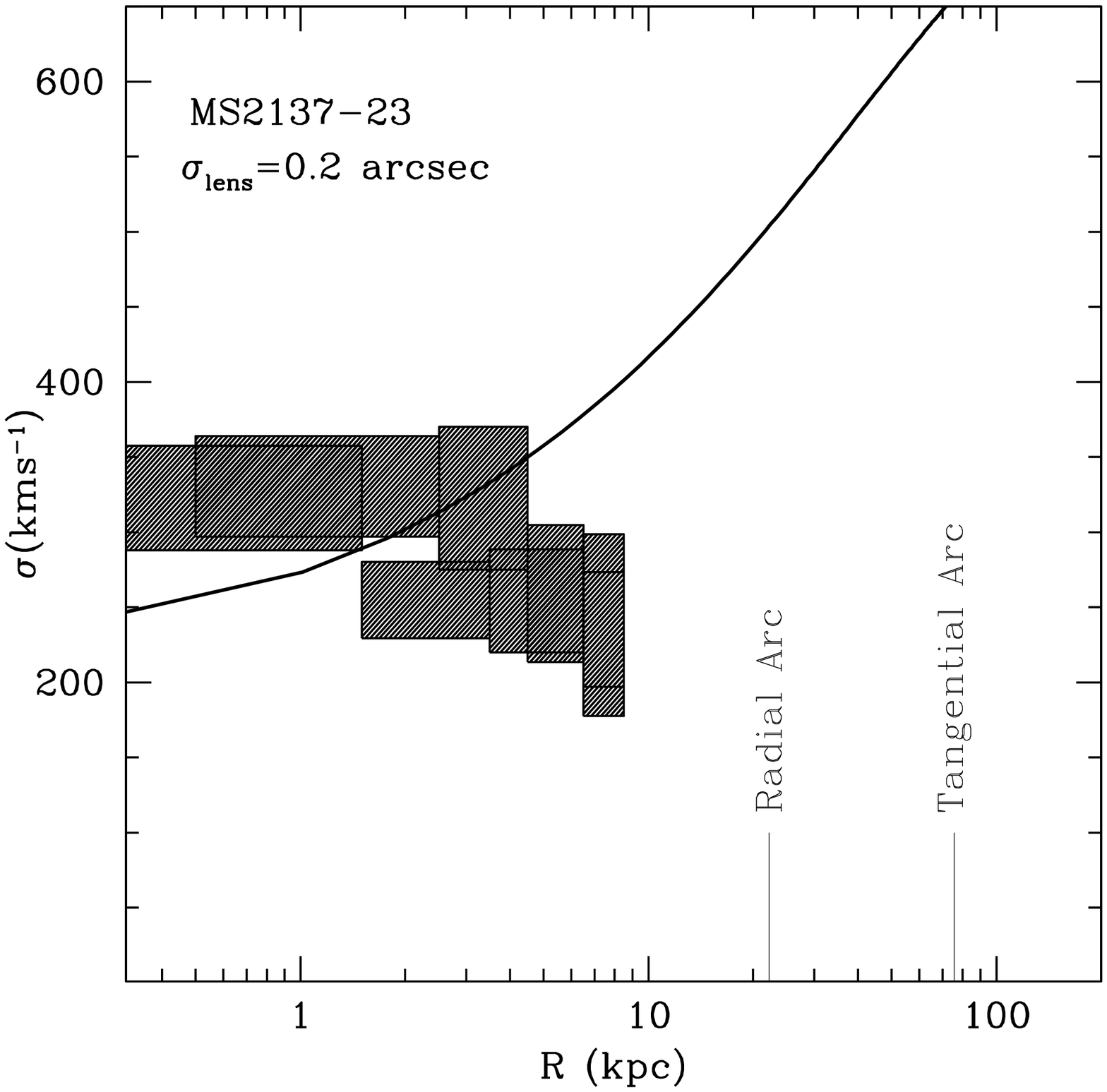}}
}
\mbox{
\mbox{\epsfysize=4.5cm \epsfbox{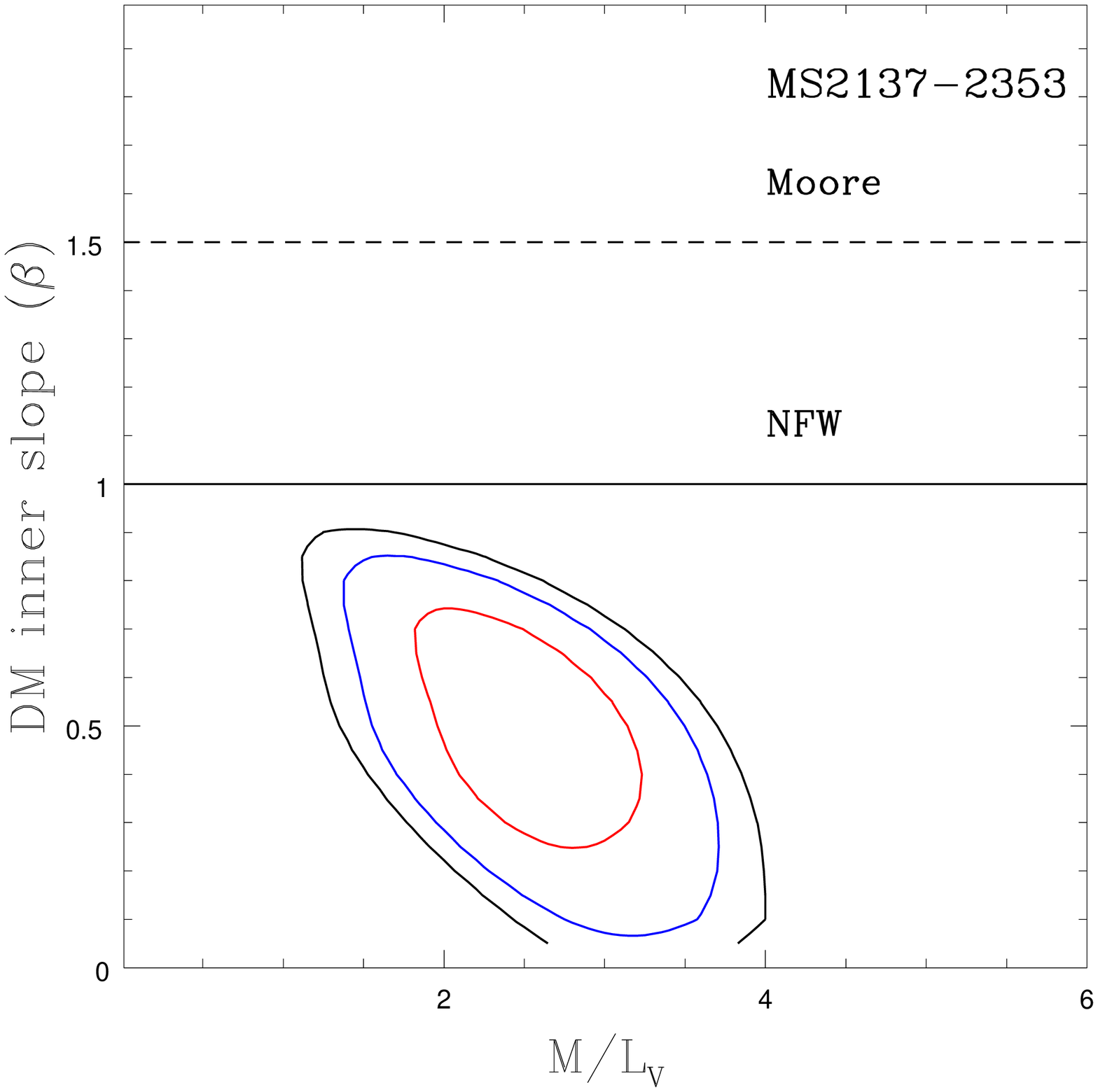}}
\mbox{\epsfysize=4.5cm \epsfbox{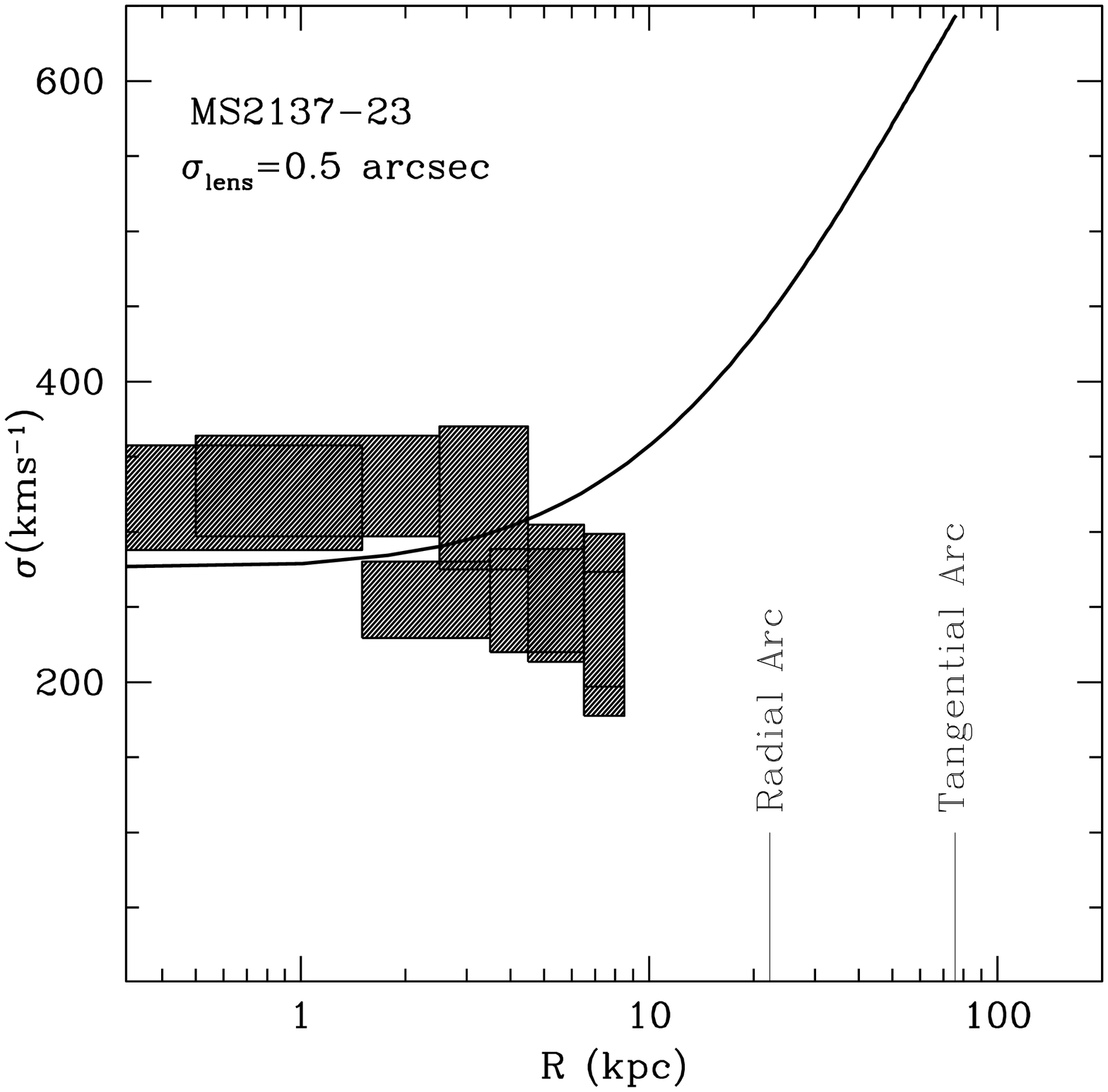}}
}
\caption{The combined lensing+dynamics results for the cluster
  MS2137-23.  The top row summarizes the results for the 0\farcs2
  lensing position uncertainty scenario while the bottom row
  encapsulates the 0\farcs5 scenario.  Top Left--Lensing+dynamics
  likelihood contours (68\%,95\%, and 99\%) in the $M/L-\beta$ plane
  after marginalizing over the other free parameters with the 0\farcs2
  lensing multiple image uncertainty.  Top Right-- Best fitting
  velocity dispersion profile from the combined lensing+dynamics
  analysis with the 0\farcs2 lensing multiple image uncertainty.  No
  models could be found that fit both the lensing and observed
  velocity dispersion constraints.  Bottom Left -- Lensing+dynamics
  likelihood contours (68\%,95\%, and 99\%) with the 0\farcs5 lensing
  multiple image uncertainty in the $M/L-\beta$ plane after
  marginalizing or optimizing over the other free parameters.  Bottom
  Right -- Best fitting velocity dispersion profile from the combined
  lensing+dynamics analysis with the 0\farcs5 lensing multiple image
  uncertainty.  While the best-fitting model velocity dispersion is a
  better fit to the data than in the 0\farcs2 lensing scenario, it
  still cannot reproduce the observed decline in the velocity
  dispersion profile in our highest radial bins -- suggesting a
  problem with our current mass model parameterization.
\label{fig:ms2137rscprior}}
\end{center}
\end{figure*}

Figure~\ref{fig:ms2137rscprior} and the discussion below summarizes
the results for the fine and coarse positional cases. One thing to
note is the number of degrees of freedom involved, i.e. the difference
between the number of constraints and the number of free parameters,
in order to quantify the goodness of fit.  The mass model has 5 free
parameters, as detailed in \S~\ref{sec:lensinterpms2137}: the DM inner
slope $\beta$, the DM pseudo-ellipticity $\epsilon$, the DM amplitude
$\delta_{c}$, the BCG stellar $M_{*}/L$, and the dark matter scale
radius $r_{sc}$ which is allowed to vary in the 100-200 kpc range.
Considering Table~\ref{tab:lensinterp}, the multiple images provide 12
constraints, while the velocity dispersion data provides 8, giving a
total of 20 data constraints.  The resulting number of degrees of
freedom is thus 15.

\subsubsection{The fine positional accuracy lensing case}

The two top panels of Figure~\ref{fig:ms2137rscprior} and the
appropriate line in Table~\ref{tab:lensmodel} encapsulate the results
of the fine positional analysis.  DM inner slopes between $\beta =
0.65 - 1.0$ lie within the 68\% confidence limit (after marginalizing
over all other free parameters), although the best fitting DM scale
radius sits at the edge of the allowed prior ($r_{sc,best}$=200 kpc).
A scale radius of 200 kpc is higher than that seen in previous lensing
analyses of MS2137-23 that have used a similar mass parameterization
to our own (e.g.\citet{gavazzi03,Gavazzi05}), so there is no case to
alter the prior.

The parameter constraints are not particularly tight because the total
$\chi^2\sim$54, larger than expected given the number of degrees of
freedom.  Such a value may indicate that the form of the mass profile
used in the fit is inappropriate. Indeed, the model velocity
dispersion profile is a poor match to that observed
(Figure~\ref{fig:ms2137rscprior}).  In fact, if the BCG velocity
dispersion results are ignored, acceptable lens models ($\chi^{2}/dof
\lesssim 1$) can be recovered with a variety of inner DM slopes, scale
radii and BCG stellar $M/L$, although these parameters have correlated
values.  We postpone discussion of the possible reasons for this
mismatch until later.

\subsubsection{The coarse positional accuracy lensing case}

The two bottom panels of Figure~\ref{fig:ms2137rscprior}, along with
Table~\ref{tab:lensmodel} summarize our results for the coarse
positional analysis.  DM inner slopes between $\beta =$0.4-0.75 lie
within our 68\% CL, and we again find DM scale radii at the high end
of our prior range, as expected.  The shift in the BCG M/L vs. DM
inner slope contour fine positional case indicates that the increased
parameter space in the lensing models has led to a slightly improved
velocity dispersion profile (bottom right panel of
Figure~\ref{fig:ms2137rscprior}). The overall $\chi^{2}\simeq$31 is
improved, although the probability for 15 degrees of freedom is less
than 1\%, assuming measurements are governed by Gaussian
statistics. Thus the model remains a marginal fit to the data.

\subsubsection{Comparison with Gavazzi 2005}

We now briefly compare our results with those of \citet{Gavazzi05}.
Gavazzi's analysis used a similar strong lensing model to our own, 
including what we consider to be somewhat less secure multiple
images. However, he extended the analysis to larger scales including
weak lensing data and incorporated the BCG velocity dispersion profile 
presented by S04.  Gavazzi adopted a strict NFW profile for the cluster 
DM halo and a Hernquist profile for the stellar component of the BCG.  

Despite these differences, Gavazzi's conclusions are very similar to
those of the present paper. Models with NFW-like DM haloes (regardless
of whether the inner slopes were varied) were uniformly poor fits to
the observational data.  In particular, the falling velocity
dispersion profile observed at $R \gtrsim 5$kpc cannot be reproduced,
despite experimenting with the effect of anisotropic orbits in the
stellar distribution.  A major conclusion of Gavazzi's study is that
halo triaxiality, an effect not typically included, may play an
important role in the central regions of galaxy clusters.  We will
return to this topic in \S~\ref{sec:ac}.

\begin{table*}
\begin{center}
\caption{Acceptable Parameter Range\label{tab:lensmodel}}
\begin{tabular}{lcccccccc}
\tableline\tableline
Prior Setup&Cluster& Inner DM Slope& $\epsilon$& $\delta_{c}$ & $r_{sc}$ & $M_*/L_{V}$ \\
&& $\beta$ & & & (kpc) &\\
\tableline
100 - 200 kpc $r_{sc}$ Prior&\\
$\sigma_{lens}=$0\farcs2& MS2137 & $0.95^{+0.05}_{-0.30}$&$0.08^{+0.01}_{-0.01}$&$29420^{+98310}_{-1760}$&$200^{-42}$&$1.58^{+0.52}_{-0.635}$\\
& Abell 383 & $0.55^{+0.2}_{-0.05}$&$0.08^{+0.01}_{-0.02}$&$140000^{+8500}_{-60600}$&$100^{+28}$& $2.4^{+0.42}_{-0.42}$\\
$\sigma_{lens}=$0\farcs5& MS2137 & $0.6^{+0.15}_{-0.2}$&$0.06^{+0.01}_{-0.01}$&$44600^{+35500}_{-7175}$&$200^{-31}$&$2.45^{+0.75}_{-0.65}$\\
& Abell 383 & $0.45^{+0.2}_{-0.25}$&$0.06^{+0.03}_{-0.01}$&$156000^{+38500}_{-67150}$&$100^{+21}$& $2.34^{+1.02}_{-0.54}$\\
\tableline
\end{tabular}
\tablecomments{Best-fitting parameters and/or confidence limits for
the different prior scenarios present in this paper.   }
\end{center}
\end{table*}

\subsection{Abell 383}

Our results for Abell 383 are shown in Figure~\ref{fig:a383rscprior}
and Table~\ref{tab:lensmodel} for both the fine and coarse positional
uncertainty cases.  We discuss each separately below.
Calculating the number of degrees of freedom in a similar way to that
done for MS2137-23, we again have 5 free parameters in our mass model.
Considering Table~\ref{tab:lensinterp}, multiple images provide 16
constraints, taking into account that those related to multiple image
sets 3,4,5 \& 6 do not have known redshift information.  The velocity
dispersion data provide 3 additional constraints.  Thus, the
resulting number of degrees of freedom is 14.

\subsubsection{The fine positional accuracy lensing case}

The top two panels of Figure~\ref{fig:a383rscprior} and the
appropriate line in Table~\ref{tab:lensmodel} summarize the results in
this case.  DM inner slopes between $\beta = 0.5 - 0.7$ lie within the
68\% confidence limit of our analysis (after marginalizing over all
other free parameters).  The best fitting scale radius sits again at
the edge of the allowed $r_{sc}$ prior range ($r_{sc}$=100 kpc).  An
X-ray analysis of Abell 383, which was able to probe to higher radius
than the current analysis, indicates that the DM scale radius is well
above 100 kpc (Zhang et al. 2007). For these reasons, and those
discussed earlier, there is no case for adjusting the DM scale radius
prior.

The total $\chi^{2}$=40.4, high given the 14 degrees of freedom in
the analysis.

\subsubsection{The coarse positional accuracy lensing case}

\begin{figure*}
\begin{center}
\mbox{
\mbox{\epsfysize=4.5cm \epsfbox{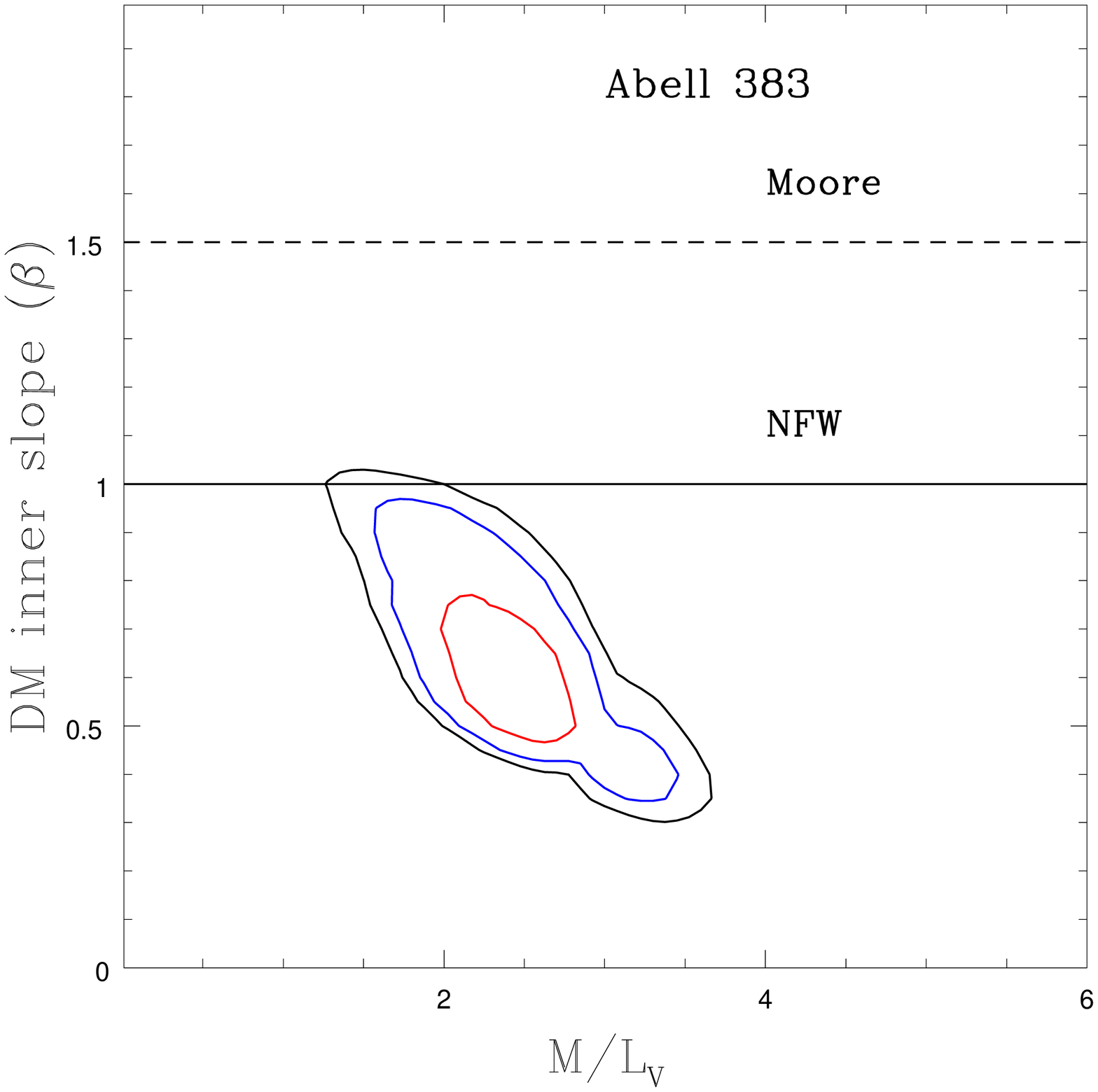}}
\mbox{\epsfysize=4.5cm \epsfbox{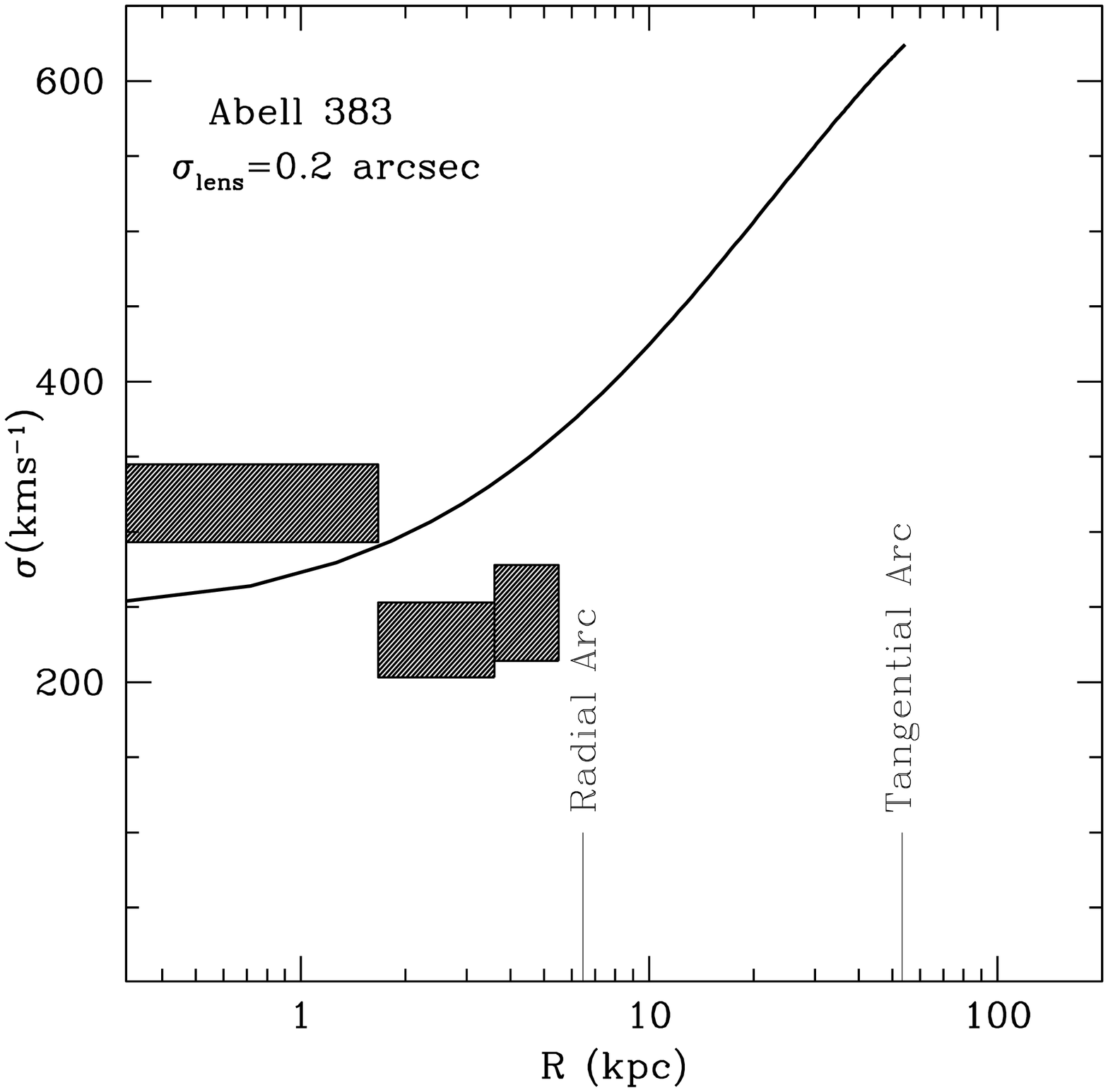}}
}
\mbox{
\mbox{\epsfysize=4.5cm \epsfbox{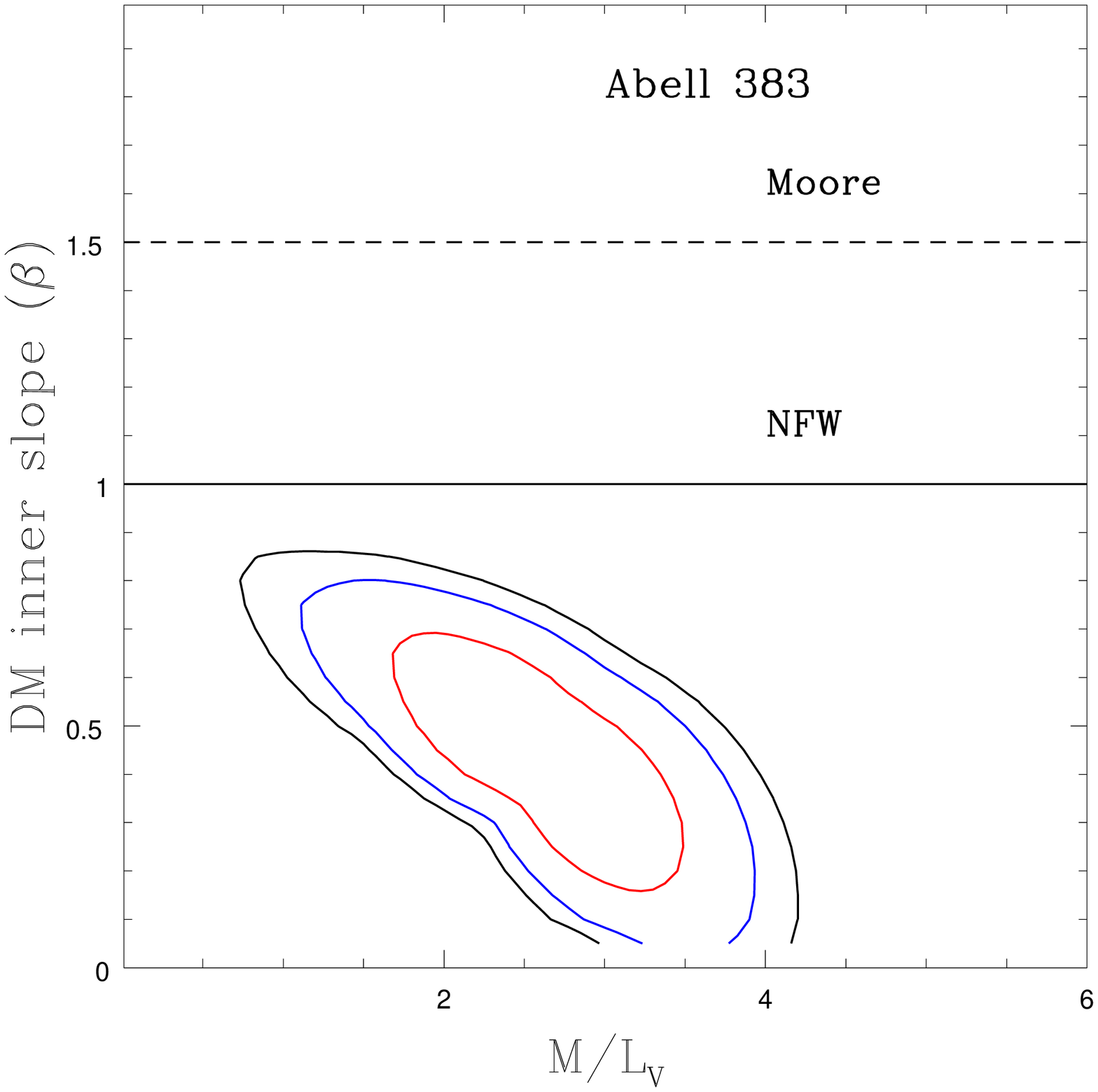}}
\mbox{\epsfysize=4.5cm \epsfbox{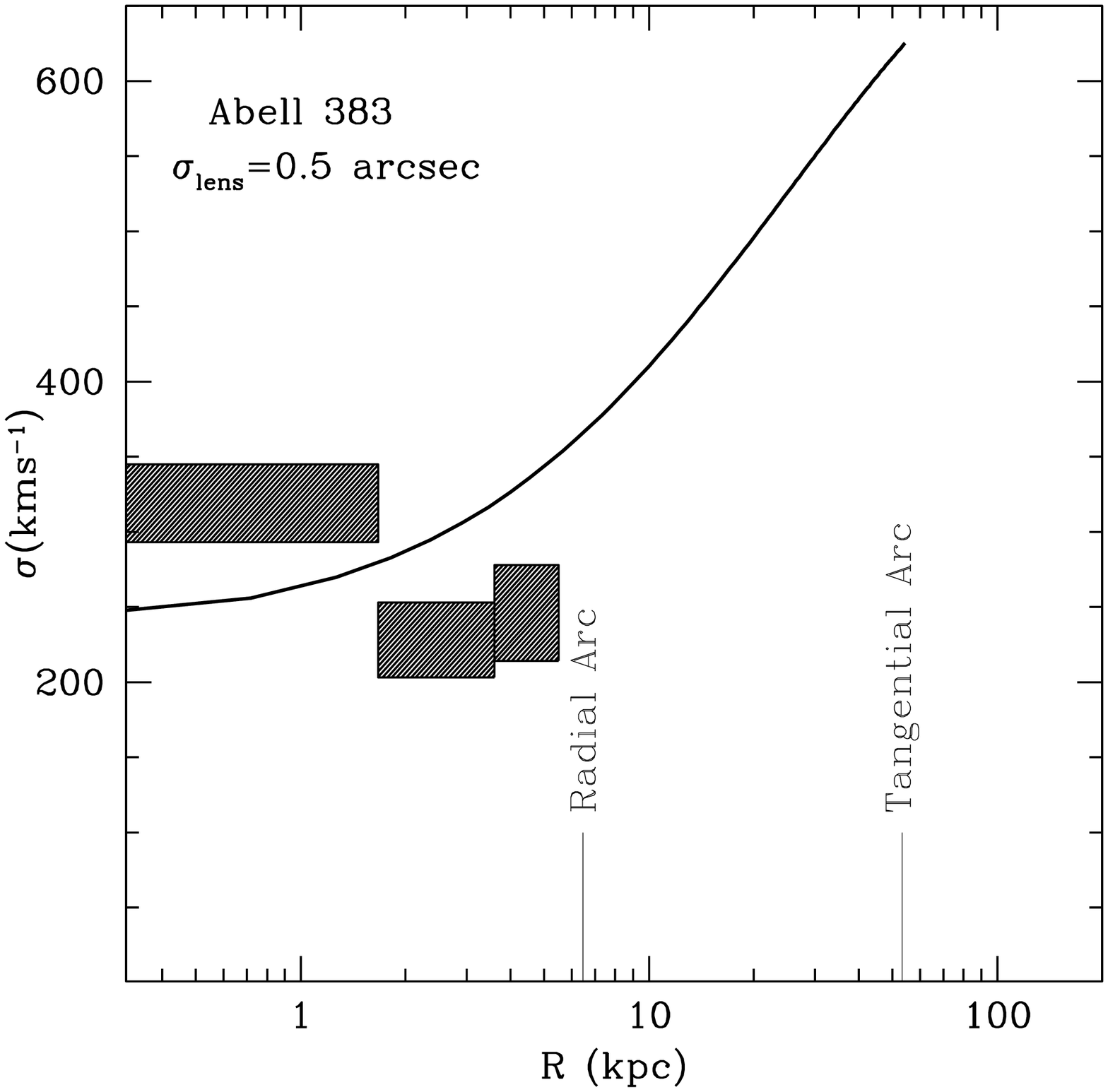}}
}
\caption{The combined lensing+dynamics results for the cluster Abell 383.
  The top row summarizes the results for the 0\farcs2 lensing position
  uncertainty scenario while the bottom row encapsulates the 0\farcs5
  scenario.  Top Left--Lensing+dynamics likelihood contours
  (68\%,95\%, and 99\%) in the $M/L-\beta$ plane with the 0\farcs2
  lensing multiple image uncertainty after marginalizing over the
  other free parameters.  Top Right-- Best fitting velocity dispersion
  profile from the combined lensing+dynamics analysis with the
  0\farcs2 lensing multiple image uncertainty.  Bottom Left --
  Lensing+dynamics contours (68\%,95\%, and 99\%) in the $M/L-\beta$
  plane with the 0\farcs5 lensing multiple image uncertainty after
  marginalization over the other free parameters.  Bottom Right --
  Best fitting velocity dispersion profile from the combined
  lensing+dynamics analysis with the 0\farcs5 lensing multiple image
  uncertainty.  The 0\farcs5 lensing multiple image scenario provides
  a better overall fit to the observations, although we are limited by
  the relatively poor quality of the observed Abell 383 velocity dispersion
  profile.
\label{fig:a383rscprior}}
\end{center}
\end{figure*}

The two bottom panels of Figure~\ref{fig:a383rscprior}, along with
Table~\ref{tab:lensmodel} summarize the results for the coarse
positional accuracy case.  DM inner slopes between $\beta =$0.2 - 0.65 
lie within our 68\% CL, along with a best fitting DM scale radius of 
100 kpc.  Our parameter constraints encompass the values found in 
the fine accuracy case with no shift in parameter space (unlike the
case for MS2137-23).  This suggests that although we should expect 
a lower $\chi^{2}$ due to the increased uncertainties allowed, no 
significant improvement to the best-fitting velocity dispersion profile 
should be expected.  As we can see in the bottom right panel of
Figure~\ref{fig:a383rscprior}, the best fitting velocity dispersion
profile is very similar to that obtained in the fine case.
The total $\chi^{2}$=22, acceptable given 14 degrees of freedom.

\section{Discussion}\label{sec:systematics}

In the previous section we have presented the results of our analysis,
which showed that a mass model comprising a stellar component for the
BCG following a Jaffe profile together with a generalized NFW DM
cluster halo is able to adequately reproduce the observations for
Abell 383 (albeit {\it only} for the coarse lensing positional
accuracy scenario) but is unable to simultaneously reproduce the
observed multiple image configuration and BCG velocity dispersion
profile for MS2137-23.  In the case of Abell 383, the inner DM profile
is flatter than $\beta$=1, supporting the earlier work of S04.  This
indicates that at least some galaxy clusters have inner DM slopes
which are shallower than those seen in numerical simulations --
but only if the mass parameterization used in the current work is
reflective of reality.  Further work in this interesting cluster using
other observational probes will further refine the mass model, 
and determine if the generalized NFW DM form is a good fit to the
cluster profile.

In this section we discuss systematic uncertainties in
our method and possible refinements that could be made to
reconcile the mass model with the observations for MS2137-23.  
We hope that many of these suggestions will become important 
as cluster mass models improve and thus will present 
fruitful avenues of research.

\subsection{Systematic Errors}

We focus first on systematic errors associated particularly with the
troublesome stellar velocity dispersion profile for MS2137-23.  Errors
could conceivably arise from (i) significant non-Gaussianity in the
absorption lines (which are fit by Gaussians), (ii) uncertain
measurement of the instrumental resolution used to calibrate the
velocity dispersion scale, and (iii) template mismatch.

Non-gaussianity introduces an error that we consider too small to
significantly alter the goodness of fit \citep{Gavazzi05}. The
instrumental resolution of ESI (the Keck II instrument used to measure
the velocity dispersion profile; \citet{Sheinis02}) is $\sim$30 \kms;
this is much smaller than the measured dispersion. Even if the
instrumental resolution was in error by a factor of two, the
systematic shift in $\sigma$ would only be 3 \kms (using Eq~3 in Treu
et al.\ 1999). This would affect all measurements and not reverse the
trend with radius.

Concerning template mismatch, S04 estimated a possible systematic
shift of up to 15-20 \kms . This could play a role especially as the
signal to noise diminishes at large radii, where the discrepancy with
the model profile is greatest. To test this hypothesis, we added 20
\kms in quadrature to only those velocity dispersion data points in
MS2137-23 at $R > 4$ kpc and recalculated the best-fitting $\chi^{2}$
values. $\chi^{2}$ is reduced from 31 to 28.8, a modest reduction
which fails to explain the poor fit.

Although selectively increasing the error bars on those data points
most discrepant with the model is somewhat contrived, our result does
highlight the need for high S/N velocity dispersion measures out to
large radii.  A high quality velocity dispersion profile has been
measured locally for Abell 2199 to $\sim$20 kpc \citep{Kelsonetal02}.
Interestingly, these high S/N measures display similar trends to those
for MS2137-23 in the overlap regime, i.e. a slightly decreasing
profile at $R\lesssim 10$kpc.  The dip witnessed in MS2137-23 is thus
not a unique feature, although with deeper measurements we might
expect to see a rise at larger radii as a result of the shallow DM
profile.
                                                                 
A final potential limitation in the dynamical analysis is the assumption
of orbital isotropy. Both S04 and \citet{Gavazzi05} explored the 
consequences of mild orbital anisotropy, concluding a possible offset 
of $\Delta \beta \sim0.15$ might result.  Even including orbital
anisotropy into his analysis, \citet{Gavazzi05} was unable to fit the 
observed velocity dispersion profile.  

Since we determine our lensing $\chi^{2}$ values in the source
plane, we checked to make sure that no extra images were seen after
remapping our best-fit lensing + velocity dispersion models back to the
image plane. No unexpected images were found, although several images
that were explicitly not used as constraints were found, such as the
mirror image of radial arc image 2a in MS2137 and the complex of
multiple images associated with 3abc, 5ab, and 6ab in Abell 383 (see
Figures~\ref{fig:mulplot} and \ref{fig:mulplota383}).  As discussed in
\S~\ref{sec:lensinterpms2137} and \ref{sec:lensinterpa383}, some of
these multiple images were not used as constraints because we could
not confidently identify their position either due to galaxy
subtraction residuals or blending with other possible multiple image
systems.  

\begin{figure*}
\begin{center}
\mbox{
\mbox{\epsfysize=4.5cm \epsfbox{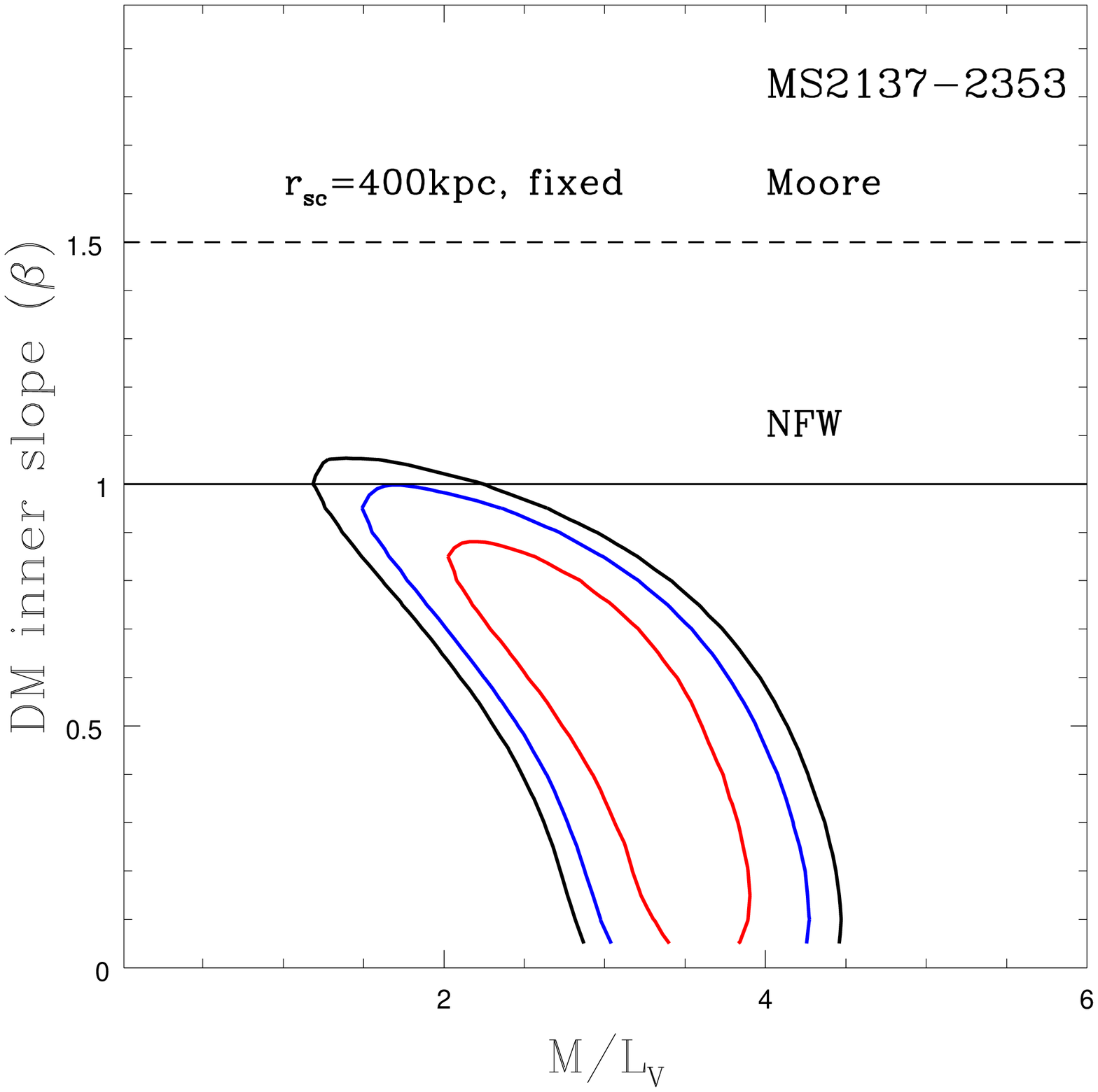}}
\mbox{\epsfysize=4.5cm \epsfbox{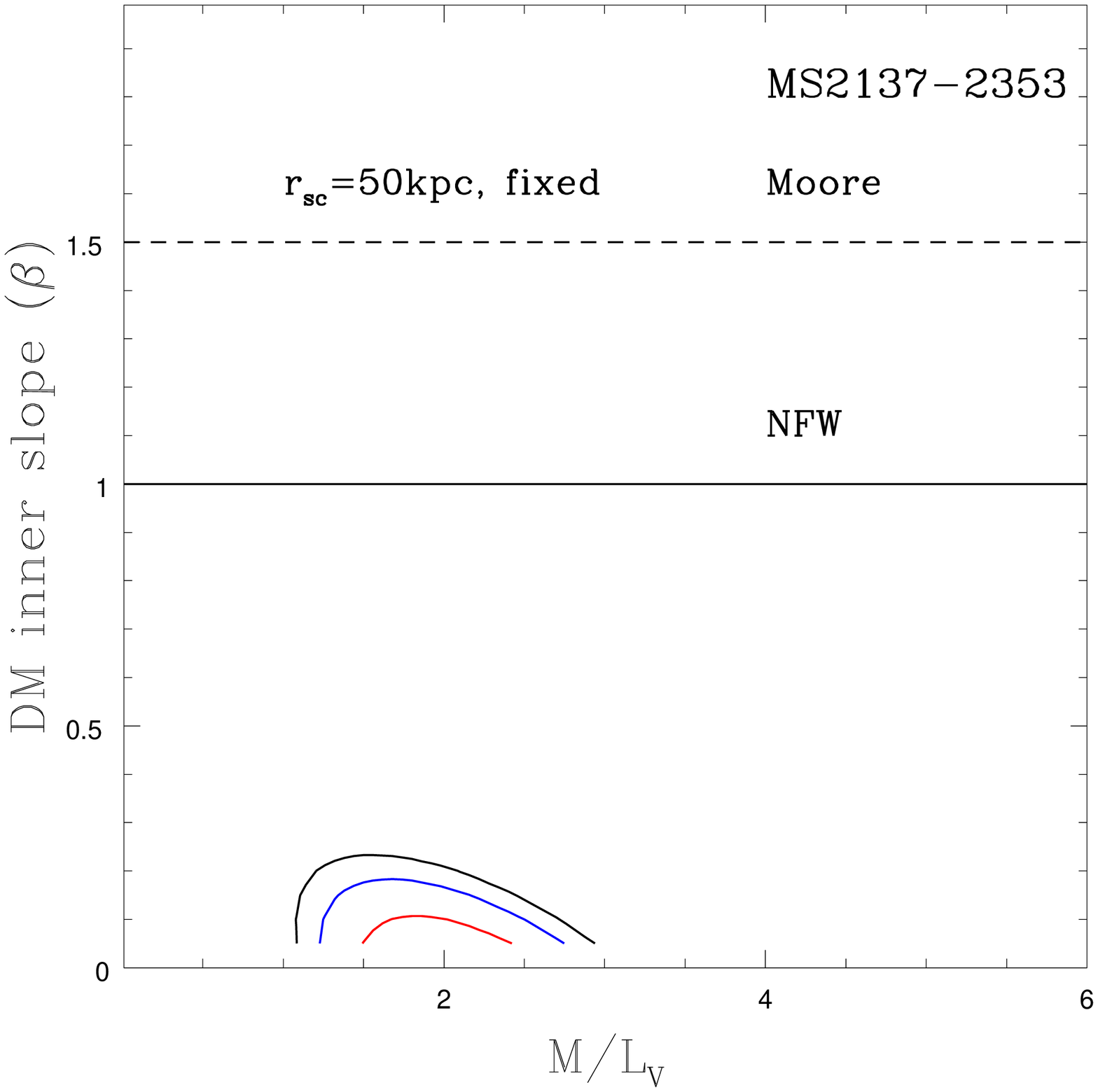}}
}
\mbox{
\mbox{\epsfysize=4.5cm \epsfbox{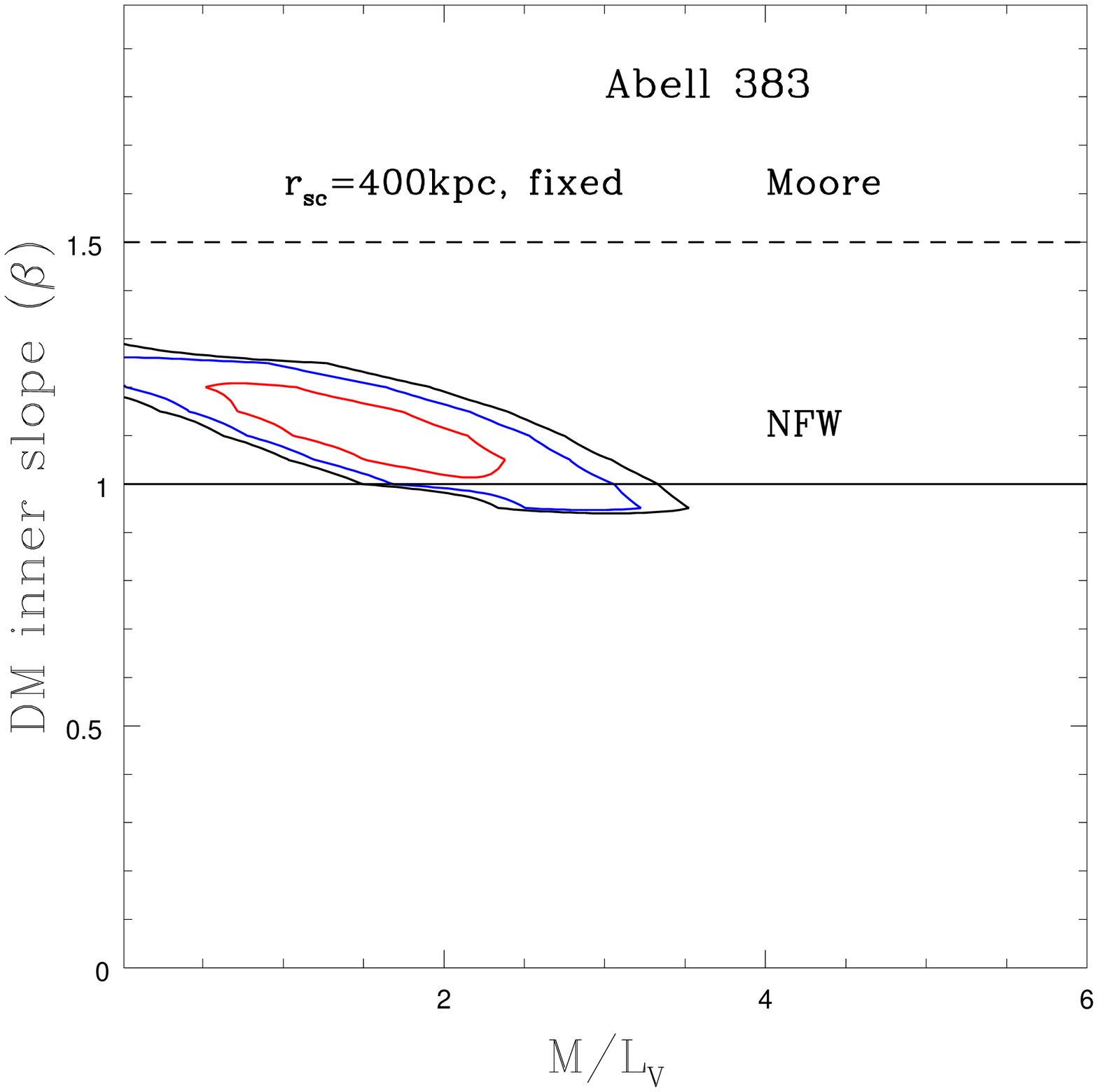}}
\mbox{\epsfysize=4.5cm \epsfbox{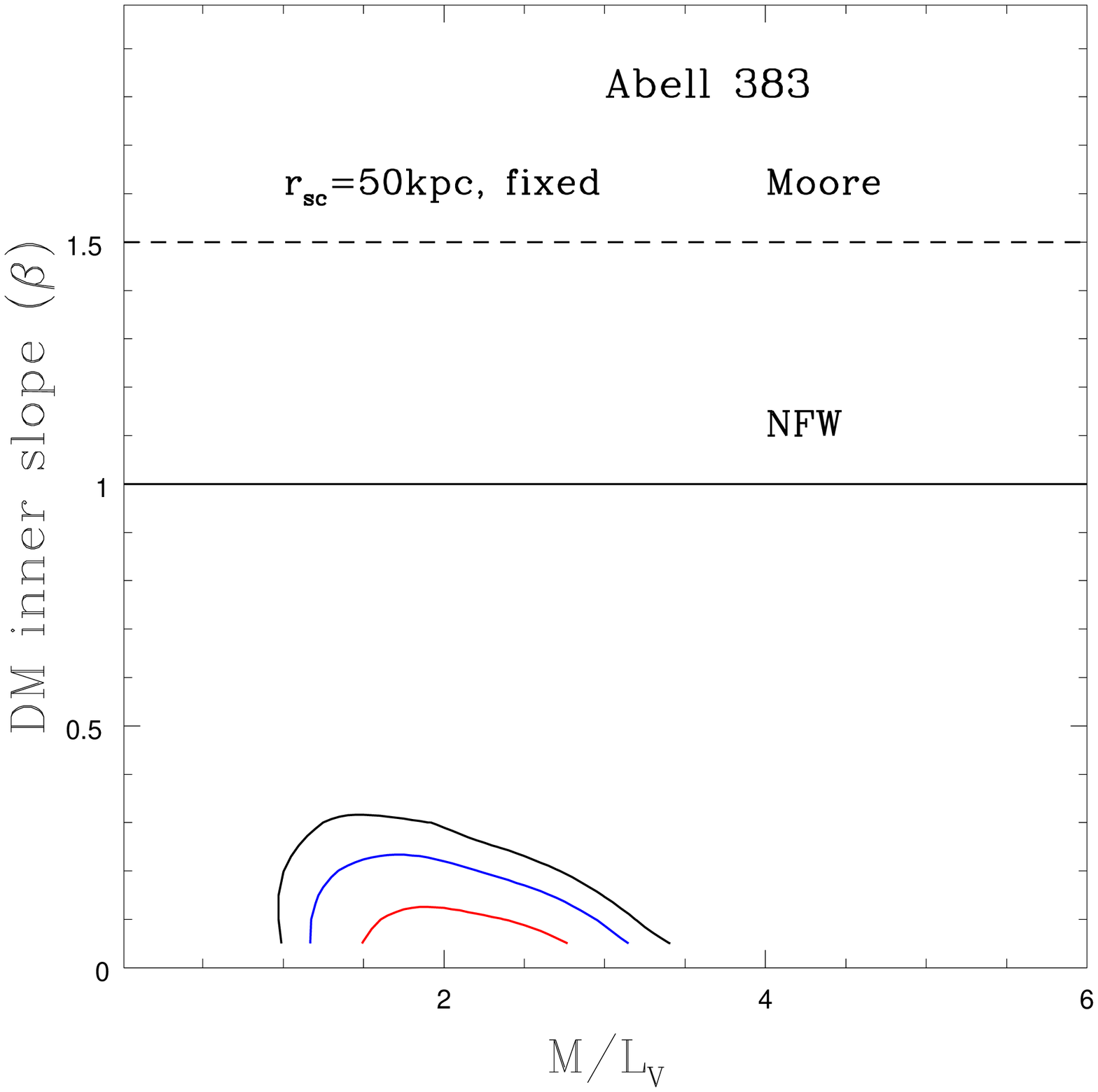}}
}
\caption{Confidence contours (68\%,95\%, and 99\%) when we allow
  the dark matter scale radius to be fixed at values a factor of two
  beyond our observationally motivated prior.  Top Row -- Contours
  when we fix the dark matter scale radius to $r_{sc}$=50 and 400 kpc
  in MS2137.  Although the $r_{sc}$=400 kpc scenario provides a
  relatively good fit to the data ($\chi^{2}\sim$26), this value for
  the scale radius is much larger than that observed from weak lensing
  data.  The $r_{sc}$=50 kpc scenario is a significantly worse fit to
  the data, with $\chi^{2}\sim$39.  Note that the DM inner slope is
  $\beta < 1$ in both scenarios.  Bottom Row -- Contours when we fix
  the dark matter scale radius to $r_{sc}$=50 and 400 kpc in A383.
  The large discrepancy in inner slope values obtained emphasize the
  need for a mass probe at larger radii.  The best-fitting model for
  either fixed scale radius is significantly worse than the
  best-fitting $r_{sc}$=100 kpc result ($\chi^{2}\sim$26.5 and 31.3
  for $r_{sc}$=50 and 400 kpc respectively).
\label{fig:diffrsc}}
\end{center}
\end{figure*}

We finally comment on the uncertainties assigned to the multiple image
systems for our lens models.  We have presented two sets of results in
this work; with assigned image positional accuracies of
$\sigma_{I}$=0\farcs2 and 0\farcs5.  We find a variety of lens models
are compatible with the $\sigma_{I}$=0\farcs2 case and only when the
velocity dispersion data is included into the analysis does the data
fail to be reproduced by the model.  Certainly if we were to further
increase the positional errors, at some point a good velocity
dispersion fit could conceivably be obtained, but we will refrain from
doing so in the present work.

Increasing the positional uncertainties is only justified if there is
evidence that there are significant missing components in the mass
models.  Further observations that can probe the mass distribution on
fine scales to larger radii and higher quality models which can
account for phenomena such as adiabatic contraction in the inner
regions of galaxy clusters and triaxiality represent the best way to
obtain a more precise picture of the cluster mass distribution.

\subsection{Improving the Mass Model}

We now turn our attention to possible inadequacies in the mass
model. It is important to stress that the two diagnostics (lensing and
dynamics) adopted in this study probe different scales.  The lensing
data tightly constrains the mass profile at and outside the radial arc
($\sim$20 kpc), while the velocity dispersion constrains the mass
profile inside $R \lesssim 10$ kpc.  Since multiple images are
numerous and their positions can be more precisely measured than
velocity dispersion \footnote{The error on the astrometry with
respect to the relevant scale, the Einstein Radius $\theta_{\rm E}$ is
much smaller than the relative error on velocity dispersion,
i.e. $\delta \theta / \theta_{\rm E} << \delta \sigma/\sigma$}, they
carry more weight in the $\chi^2$ statistic than the kinematic points,
producing a best overall fitting model (which is lensing dominated)
that is a relatively poor fit to the kinematic data.  To improve the
model, one must admit that either one of the two components of the
modeling is incorrect, or that the functional form of the mass profile
chosen to extrapolate the lensing information at the scales relevant
for dynamics is insufficient.  In this section we discuss several
areas where the mass model presented in this paper could be improved.

\subsubsection{The Contribution of the Brightest Cluster Galaxy}

We might query the assumption of a Jaffe density profile for the BCG.  
This seems an unlikely avenue for improvement given the Jaffe profile 
fits the observed BCG surface brightness profile remarkably well 
(see Figure 2 of S04).  Moreover, \citet{Gavazzi05} utilized a Hernquist 
mass profile in his analysis of MS2137-23, which also matches the 
observations, and \citet{Gavazzi05} was likewise unable to reproduce 
the observed S04 velocity dispersion profile.

We have additionally checked our assumptions by altering the
PIEMD fit to the BCG surface brightness data so that it is matched not
to the derived Jaffe profile fit to the BCG but directly to the HST
surface brightness profile.  With this setup, we found a $r_{cut}$ value
of 23.70 kpc for MS2137 and 28.65 kpc for Abell 383 (compare this with
the numbers in Table~\ref{tab:lensfixed}).  Redoing our analysis for
the best-fitting $r_{sc}$ scenario only, our constraints on $\beta$
for both Abell 383 and MS2137 did not change by more than 0.05, and so
it is not likely that our method for constraining the BCG mass
contribution is the root cause of our inability to fit the data to a
BCG + gNFW cluster DM halo mass model.

Conceivably the BCG may not be coincident with the center of the
cluster DM halo, as has been assumed throughout this work.  It is
often the case that small subarcsecond off sets between BCGs and
cluster DM halos are necessary to fit lensing constraints
(e.g.~\citet{gps05}).  There is strong evidence that the BCG is
nearly coincident with the general cluster DM halo {\it in projection}
from the strong lensing work presented here and by others
\citep{gavazzi03,Gavazzi05}.  However, an offset could be responsible
for the flat to falling observed velocity dispersion profile if the
BCG were actually in a less dark matter dominated portion of the
cluster.  Another possibility is that there are multiple massive
structures along the line of sight, which would be probed by the
strong lensing analysis, but not with the velocity dispersion profile
of the BCG.  A comprehensive redshift survey of MS2137-23 could
provide further information on structures along the line of sight.

\subsubsection{The Advantage of a Mass Probe at Larger Radii}\label{sec:highrad}

With our presented data set, we have seen that it is difficult to
constrain the DM scale radius, $r_{sc}$ because both of our mass
probes are only effective within the central $\sim$100 kpc of the
clusters -- within the typical DM scale radius observed and seen in
CDM simulations.  For this reason, the inferred DM scale radius for
both Abell 383 and MS2137-23 lay at the boundary of our assumed prior
range.  Future work will benefit from weak lensing data, along with
galaxy kinematics and X-ray data of the hot ICM which can each probe
out to large clustercentric radii.

Although not the focus of the current work, pinning down the correct
DM scale radius will be crucial for constraining other DM mass
parameters.  For instance, there is a well-known degeneracy between
$r_{sc}$ and the inner slope $\beta$
(e.g.~\citet{gavazzi03,Gavazzi05}). To briefly explore this, we
have reran our analysis (for the coarse positioning lensing case) for
both clusters with a $r_{sc}$ of 50 and 400 kpc -- factors of two
beyond our chosen $r_{sc}$ prior.  We show our confidence contours in
Figure~\ref{fig:diffrsc}, which are noteworthy.  For example in the
case of MS2137-23, if we fix $r_{sc}$=50 kpc, then the best-fitting
$\beta = 0.05$.  However, if $r_{sc}$=400 kpc then $\beta=0.7$, more
in accordance with simulations.  Interestingly, the $r_{sc}$=400kpc
scenario returns a better overall $\chi^{2}\sim26$ than any model with
$r_{sc}$=100-200 kpc -- even though a $r_{sc}$ of 400 kpc is clearly
ruled our by extant weak lensing observations.  None of the other
$r_{sc}$=50,400 kpc scenarios produced $\chi^{2}$ values that were
comparable to those seen with $r_{sc}$=100-200 kpc.  Any further
knowledge of the DM scale radius would aid greatly in constraining
$\beta$ and determining the overall goodness of fit of the generalized
NFW DM profile to the cluster data.

X-ray studies assuming hydrostatic equilibrium
\citep{Allen01vir,Schmidt06} and a combined strong and weak lensing
analysis \citep{Gavazzi05} have presented data on MS2137-23 to radii
much larger than that probed in this study. To check that the mass
model derived from data within $\sim$ 100 kpc do not lead to results
at variance with published data at larger radii, we have taken the
\citet{Gavazzi05} results and compared their derived mass at large
radii with an extrapolation of our mass models.

Examining Figure~3 from \citet{Gavazzi05} we estimate that from his
weak lensing analysis a 2D projected mass enclosed between $1.6 \times
10^{14}$ and $1.1 \times 10^{15} M_{\odot}$ at $\sim 1.08$ Mpc using
the cosmology adopted in this paper.  Correspondingly, if we take all
of the $\Delta \chi^{2}<1.0$ models using our analysis method (the
coarse positional accuracy case was use) and calculate the expected 2D
projected mass enclosed at 1.08 Mpc we find values between $6.9 \times
10^{14}$ and $8.4 \times 10^{14} M_{\odot}$, well within the expected
range.

Note that no attempt was made to extrapolate the mass {\it profiles}
derived in our analysis to larger radii than the data in this paper
allow, although we are acquring weak lensing data for a large
sample of galaxy clusters to perform a more extensive analysis. The
purpose of this consistency check is to only ensure that the masses we
derive for such large radii are not too discrepant with existing
analyses.  The consistency check is satisfied and lends some credence
to the models.

\subsubsection{Dark matter baryons interactions and Triaxiality}\label{sec:ac}

The central regions of DM halos can be strongly affected by the
gravitational interaction with baryons during halo formation.  If
stars form and condense much earlier than the DM, it is expected that
the baryons will adiabatically compress the DM resulting in a halo
that is {\it steeper} than that of the original
\citep{Blumenthal86,Gnedin04}.  Alternatively, dark matter heating
through dynamical friction with cluster galaxies can counteract
adiabatic contraction, leading to a shallower DM profile
\citep{Elzant04,Nipoti03}.  The present work takes into account
neither of the above scenarios, and if any baryon-DM interaction
greatly changes the cluster density profile, our assumed parameterized
gNFW profile may be inappropriate.  Recently, \citet{Zappacosta06}
have used X-ray mass measurements in the cluster Abell 2589 to
conclude that processes in galaxy cluster formation serve to
counteract adiabatic contraction in the cluster environment.
Certainly, more observational work is needed to understand the
interplay between baryons and DM in clusters, and extended velocity
dispersion profiles of BCGs in conjunction with other mass tracers at
larger radii could serve as the best testing ground for the interplay
of dark and luminous matter.

Not only is there likely significant interplay between baryons and DM
in the central regions of clusters, but real galaxy clusters are
certainly triaxial and, if ignored, this may lead to biased parameter
estimations and discrepancies when combining mass measurement
techniques that are a combination of two- and three-dimensional.
Several recent studies have considered the effects of halo triaxiality
on observations.  Using an N-body hydrodynamical simulation of a disk
galaxy and performing a 'long slit' rotation curve observation,
\citet{Hayashi04} found that orientation and triaxial effects can
mistake a cuspy DM profile for one that has a constant density core.
At the galaxy cluster scale, \citet{Clowe04} performed mock weak
lensing observations of simulated galaxy clusters and found that the
NFW concentration parameter recovered was correlated with the 3D
galaxy cluster orientation.  In order to investigate the recent rash
of galaxy clusters with observed high concentration parameters in
seeming contradiction to the CDM paradigm
\citep{Kneib03,Gavazzi05,Broadhurst05b}, \citet{Oguri05} used strong
and weak lensing data in Abell 1689 along with a set of models that
included halo triaxiality and projection effects.  Again, it was seen
that halo shape causes a bias in mass (and mass profile)
determination, although it should be kept in mind that measurements of
concentration are extremely difficult (e.g. Halkola et al.\ 2006), and
the recent study of \citet{Limousin06} has seemed to clear up the
concentration parameter controversy for at least Abell 1689.

In terms of the current work, \citet{Gavazzi05} has pointed out that
the inability of his lensing model to fit the MS2137-23 BCG velocity
dispersion profile may be due to halo triaxiality or projected mass
along the line of sight (which would increase the mass measured in the
lensing analysis but would not be seen in the stellar velocity
dispersion).  \citet{Gavazzi05} showed that an idealized prolate halo
with an axis ratio of $\sim$ 0.4 could explain the velocity dispersion
profile in MS2137-23.  Halo triaxiality could also explain the high
concentration previously seen in this cluster.  Again, the gap between
simulations and observations may be bridged with respect to
triaxiality if further steps were taken to compare the two directly.
One step in this direction would be the publication of detailed
density profiles for the simulations (in 3-D or along numerous
projected sight-lines).

The most recent DM only simulations have indicated that the
standard NFW profile representation of a DM profile (and its
\citet{M99} counterpart with an inner slope $\beta \sim 1.5$) can be
significantly improved by slightly altering the model to a profile
with a slope that becomes systematically shallower at small radii
(e.g.~\citet{Navarro04}, but see \citet{Diemand05}).  While we have
adopted the traditional generalized NFW profile in this study, future
work with parameterized models should move towards the latest fitting
functions along with an implementation of adiabatic contraction as has
already been attempted by \citet{Zappacosta06}.  Note, however, that
both \citet{Navarro04} and \citet{Diemand04} have stated that all
fitted functions have their weaknesses when describing complicated
N-body simulations and that when possible simulations and observations
should be compared directly.

\section{Summary \& Future Work}\label{sec:finale}

We have performed a joint gravitational lensing and dynamical analysis
in the inner regions of the galaxy clusters Abell 383 and MS2137-23 in
order to separate luminous baryonic from dark matter in the cluster
core.  To achieve this, we implemented a new 2D pseudo-elliptical
generalized NFW mass model in an updated version of the {\sc LENSTOOL}
software package. This refinement is a natural progression from
our earlier attempts to measure the dark matter density profile
\citep{Sand02, Sand04}.

For the study, we adopted an observationally motivated scale radius
prior of $r_{sc}=100-200$ kpc.  With strong lensing alone, we find
that a range of mass parameters and DM inner slopes are compatible
with the multiple image data, including those with $\beta > 1$ as seen
in CDM simulations.  However, including the BCG kinematic constraints
for both systems, the acceptable parameter ranges shrink
significantly.

We can summarize the results for the two clusters as follows:

\begin{enumerate}

\item For the cluster Abell 383 we have found satisfactory BCG +
generalized NFW cluster DM models only for our coarse lensing
positional accuracy scenario.  Assuming that this is reflective of the
underlying cluster DM distribution, the dark matter inner slope is
found to be $\beta=0.45^{0.2}_{-0.25}$, supporting our earlier
contention that some clusters have inner DM profiles flatter than
those predicted in numerical simulations.

\item For MS2137-23 our model is unable to reproduce the observed BCG
velocity dispersion profile and the range of accepted inner slopes
therefore depends sensitively on the adopted uncertainties in the mass
model.  This may suggest an unknown systematic uncertainty in our
analysis or that we have adopted an inappropriate mass model.  We
explore the former in considerable detail, extending the quite
extensive discussion of \citet{Sand04}.  However, no obvious cause can
be found. If, as we suspect, the cause lies with our adopted mass
model, it points to the need for further work concerning the
distribution of dark matter in the central regions of galaxy
clusters.

\end{enumerate}

Future modeling efforts should include the effects of triaxiality and
the influence of baryons on dark matter.  It is also critical to
obtain high S/N extended velocity dispersion measurements of more BCGs
out to larger radii so that, in conjunction with other mass
measurement techniques, the interplay of baryons and dark matter in
cluster cores can be studied with a real sample.  Some other future
directions are straightforward.  For example, the deep multiband ACS
imaging now being done with galaxy clusters \citep{Broadhurst05} allow
for literally hundreds of multiple images to be found, significantly
increasing the number of constraints and allowing for nonparametric
mass modeling \citep{Diego04} -- a crucial addition in case the
currently used parameterized models do not correspond to reality.  We
are eager to find ways to more directly compare simulations with
observations so that clearer conclusions can be drawn over whether or
not simulations and observations are compatible.  This may
involve measuring other properties of the dark matter halo rather than
a sole emphasis on the inner slope, such as the concentration
parameter, $c$.  Simulated observations of numerical simulations,
such as that presented recently by \citet{Meneghetti05}, offer a clear
way forward in understanding the systematics involved in observational
techniques and the kinds of observations required to test the current
paradigm for structure formation.

\acknowledgements

We thank Raphael Gavazzi for numerous stimulating conversations. DJS
acknowledges support provided by NASA through Chandra Postdoctoral
Fellowship grant number PF5-60041.  TT acknowledges support from the
Sloan Foundation through a Sloan Research Fellowship. GPS acknowledges
financial support from a Royal Society University Research Fellowship.
Finally, the authors wish to recognize and acknowledge the cultural
role and reverence that the summit of Mauna Kea has always had within
the indigenous Hawaiian community.  We are most fortunate to have the
opportunity to conduct observations from this mountain.  This research
has made use of the NASA/IPAC Extragalactic Database (NED) which is
operated by the Jet Propulsion Laboratory, California Institute of
Technology, under contract with the National Aeronautics and Space
Administration.

\appendix

\section{A Generalized NFW Implementation in LENSTOOL}

Here we briefly discuss the implementation of the pseudo-elliptical
generalized NFW profile into the {\sc lenstool} software package.  The
interested reader is referred to \citet{Kneibphd,gps05} for further
details about the {\sc lenstool} software.  Some of what follows has
been presented by \citet{Golse02}, but is reviewed here for continuity
and clarity.

Throughout this section we are using the thin-lens approximation with
$r^{2}=R^{2}+z^{2}$ and $\bf{x} = (x_{1},x_{2})= \bf{R} /r_{sc}$.  By
introducing ellipticity into the potential rather than the surface
mass density we make the lensing calculations more tractable given
that the deflection angle is just the gradient of the scaled lensing
potential.  Using the following coordinate substitution of $x$ by
$x_{\epsilon}$,

\begin{equation}
\label{defin_ell}
\left\lbrace
\begin{array}{lcl}
x_{1\epsilon} & = & \sqrt{a_{1}} \, x_1 \\
x_{2\epsilon} & = & \sqrt{a_{2}} \, x_2 \\
x_\epsilon & = & \sqrt{x_{1\epsilon}^2 +
x_{2\epsilon}^2}\ =\  \sqrt{a_{1}x_1^2 +a_{2}x_2^2}\\
\phi_\epsilon & = & \arctan \left(x_{2} / x_{1}\right)
\end{array}
\right.
\end{equation}

\noindent where $a_{1}$ and $a_{2}$ are two elliptical parameters.  We
can calculate the elliptical deflection angle:

\begin{equation}
\vec{\alpha}_\epsilon(\vec{x})=\left(
\begin{array}{l}
\displaystyle{\frac
{\partial\varphi_\epsilon}{\partial x_1}}=
\alpha(x_\epsilon)\sqrt{a_{1}}\cos{\phi_\epsilon}\\
\displaystyle{\frac
{\partial\varphi_\epsilon}{\partial x_2}}=
\alpha(x_\epsilon)\sqrt{a_{2}}\sin{\phi_\epsilon}\\
\end{array}
\right)
\label{defl_ell}
\end{equation}
The above expression holds for any definition of $a_{1}$ and $a_{2}$
which we choose to be:

\begin{equation}
\label{a_GK}
\begin{array}{c}
a_{1}=1+\epsilon\\
a_{2}=1-\epsilon
\end{array}
\end{equation}

\noindent While this choice of $a_{1}$ and $a_{2}$ do not correspond
directly to the ellipticity of the potential (see
\citealt{Meneghetti03b}, who use a different parameterization), it
does lead to simple expressions for standard lensing quantities, such
as the surface mass density and shear (see Eqs. 17-19 of
\citet{Golse02}) .  The standard ellipticity of the potential
$\epsilon_{\varphi}$ is related to $\epsilon$ by

\begin{equation}
\label{eq:phi}
\epsilon_{\varphi} = 1 - \displaystyle{\sqrt{\frac{1-\epsilon}{1+\epsilon}}}.
\end{equation}

Using the standard lensing functions (see e.g.~\citet{Miralda91}) for
the deflection angle ($\alpha$), convergence ($\kappa$), and shear
($\gamma$), along with Eqn~\ref{defl_ell} above, the projected mass
density $\Sigma_\epsilon(\vec{x})$ for our pseudo-elliptical
implementation is simply:

\begin{equation}
\Sigma_\epsilon(\vec{x})=\Sigma(\vec{x}_\epsilon)+\epsilon\cos{2\phi_\epsilon}
(\overline{\Sigma}(\vec{x}_\epsilon)-\Sigma(\vec{x}_\epsilon)).
\label{sigma_ell}
\end{equation}

Likewise, the 3D pseudo-elliptical density profile can be similarly
derived to be

\begin{equation}
\rho_\epsilon(\vec{x},z)=\rho(\vec{x}_\epsilon,z)+\epsilon\cos{2\phi_\epsilon}
(\frac{2}{x_{\epsilon}^{2}} \int_{0}^{x_{\epsilon}}x \rho(\vec{x},z) dx-\rho(\vec{x}_\epsilon,z))
\label{rho_ell}
\end{equation}

\noindent where $z$ is the direction along the line of sight.  Since
we are not making an effort to probe the triaxiality of our galaxy
clusters, we will plot projected quantities whenever possible.

Although Eqns~\ref{rho_ell} and \ref{sigma_ell} are general, we are
working with the generalized NFW density profile

\begin{equation}
\label{eq:pgnfw}
\rho_d(r)=\frac{\rho_{c} \delta_{c}}{(r/r_{sc})^{\beta}(1+(r/r_{sc}))^{3-\beta}}
\end{equation}

and the resulting surface density profile

\begin{equation}
\label{eq:sddm}
\Sigma_{gNFW} (R)=2 \rho_{c} r_{sc} \delta_{c} x^{1-\beta} \int_0^{\pi/2}d\theta \sin\theta (\sin\theta + x)^{\beta-3}
\end{equation}
when modeling the cluster dark matter halos.

\noindent In Figure~\ref{fig:ellip_ill} the extent to which our
implementation does not produce surface density profiles with true
elliptical isocontours is illustrated.  As the parameter $\epsilon$
increases the surface density isocontours become more boxy and peanut
shaped.  However, at relatively low $\epsilon$, the isocontours are
very nearly elliptical.  We discuss in the following section to what
extent our generalized NFW pseudo-elliptical mass model is an adequate
description of an elliptical mass distribution.

Unlike the NFW profile, the surface mass density and
deflection angle of the generalized NFW profile cannot be calculated
analytically.  This greatly slows any lensing computation, especially
when we need to calculate $\chi^{2}$ values over large parameter
hypercubes.  To limit the computing time necessary, we created a
look-up table for all of the necessary integrals from which we
interpolate when performing our lensing calculations.

\subsection{Limitations of the Pseudo-Elliptical Treament}

\begin{figure*}
\begin{center}
\mbox{\epsfysize=6.0cm \epsfbox{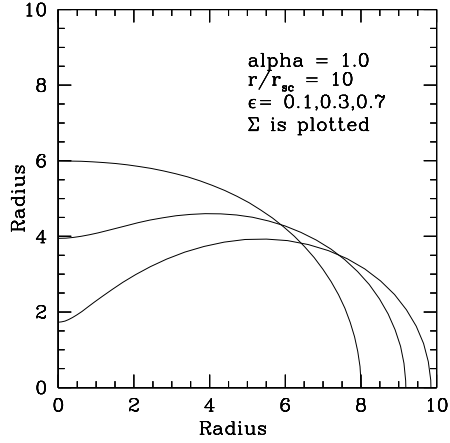}\label{fig:ellip_ill}}
\caption{Illustration of projected density isocontours for the
pseudo-elliptical generalized NFW parameterization with
$r/r_{sc}=10.0$ and $\beta=1.0$.  Note that as $\epsilon$ gets larger,
the projected density contours become more dumb-bell shaped.}

\end{center}
\end{figure*}

\begin{figure*}
\begin{center}
\mbox{
\mbox{\epsfysize=5.5cm \epsfbox{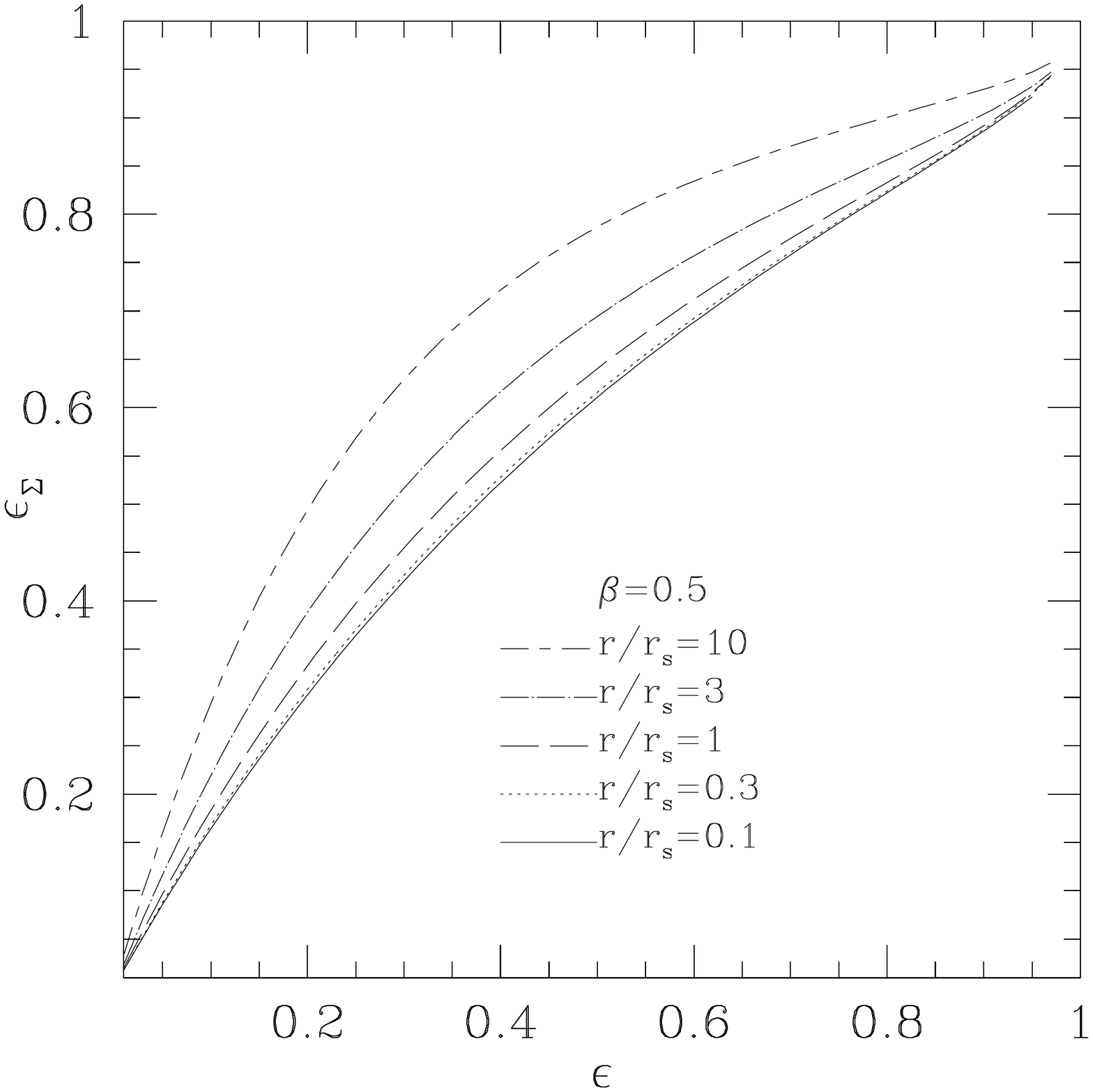}}
\mbox{\epsfysize=5.5cm \epsfbox{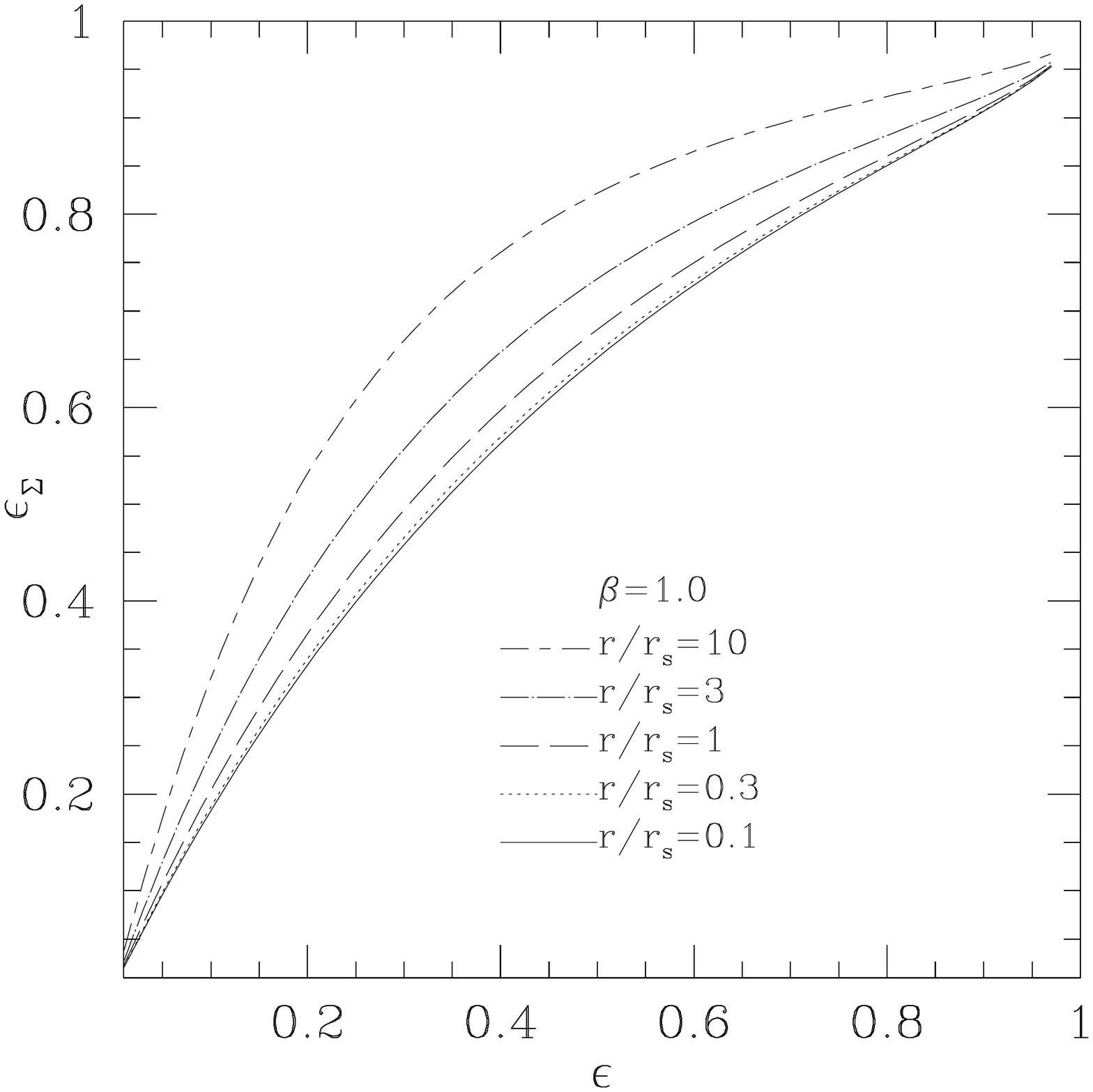}}
\mbox{\epsfysize=5.5cm \epsfbox{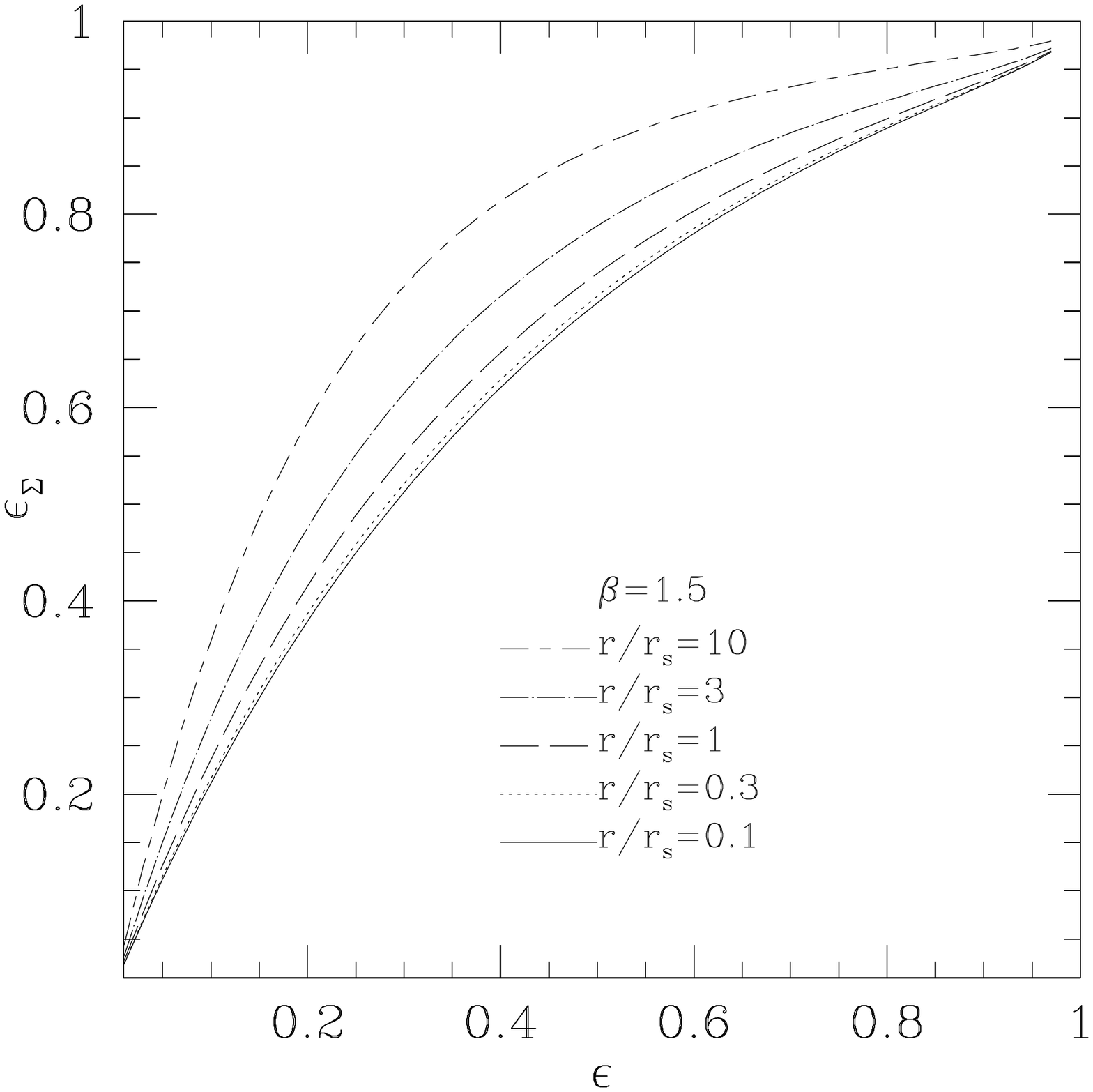}}
}
\caption{The ellipticity (taking the minor and major axis positions
and assuming $\epsilon_{\Sigma}=1-b/a$) of the projected density,
$\Sigma$, as a function of the ellipticity in the potential for
different values of $\beta$.
\label{fig:evsepot}}
\end{center}
\end{figure*}

In this section we quantitatively investigate the range of $\epsilon$
for which the generalized NFW pseudo-elliptical mass model is an
adequate description of an elliptical mass distribution.  As can be
seen in Figure~\ref{fig:ellip_ill} our pseudo-elliptical
representation can depart strongly from a true elliptical model at
high ellipticities.  To what degree can we consider out treatment of
ellipticity an accurate one for representing elliptical surface
density distributions?  To answer this, we reapply several of the
quantitative measures presented by \citet{Golse02} to our generalized
NFW pseudo-elliptical model.  To get a feel for the relation between
$\epsilon$ and the ellipticity in the surface mass density,
$\epsilon_{\Sigma}$, we plot several values in Fig.~\ref{fig:evsepot}.

\begin{figure*}
\begin{center}
\mbox{
\mbox{\epsfysize=5.5cm \epsfbox{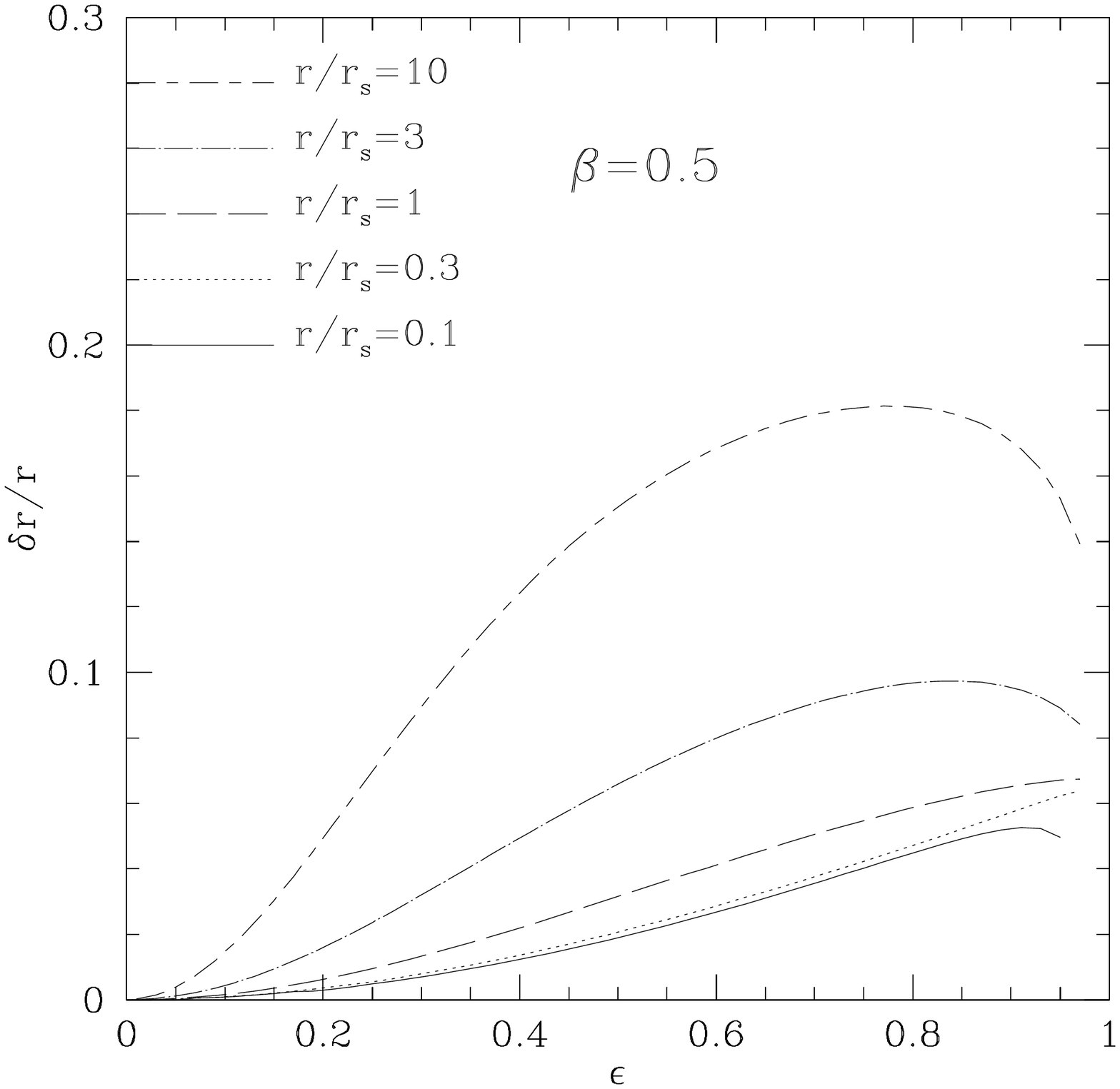}}
\mbox{\epsfysize=5.5cm \epsfbox{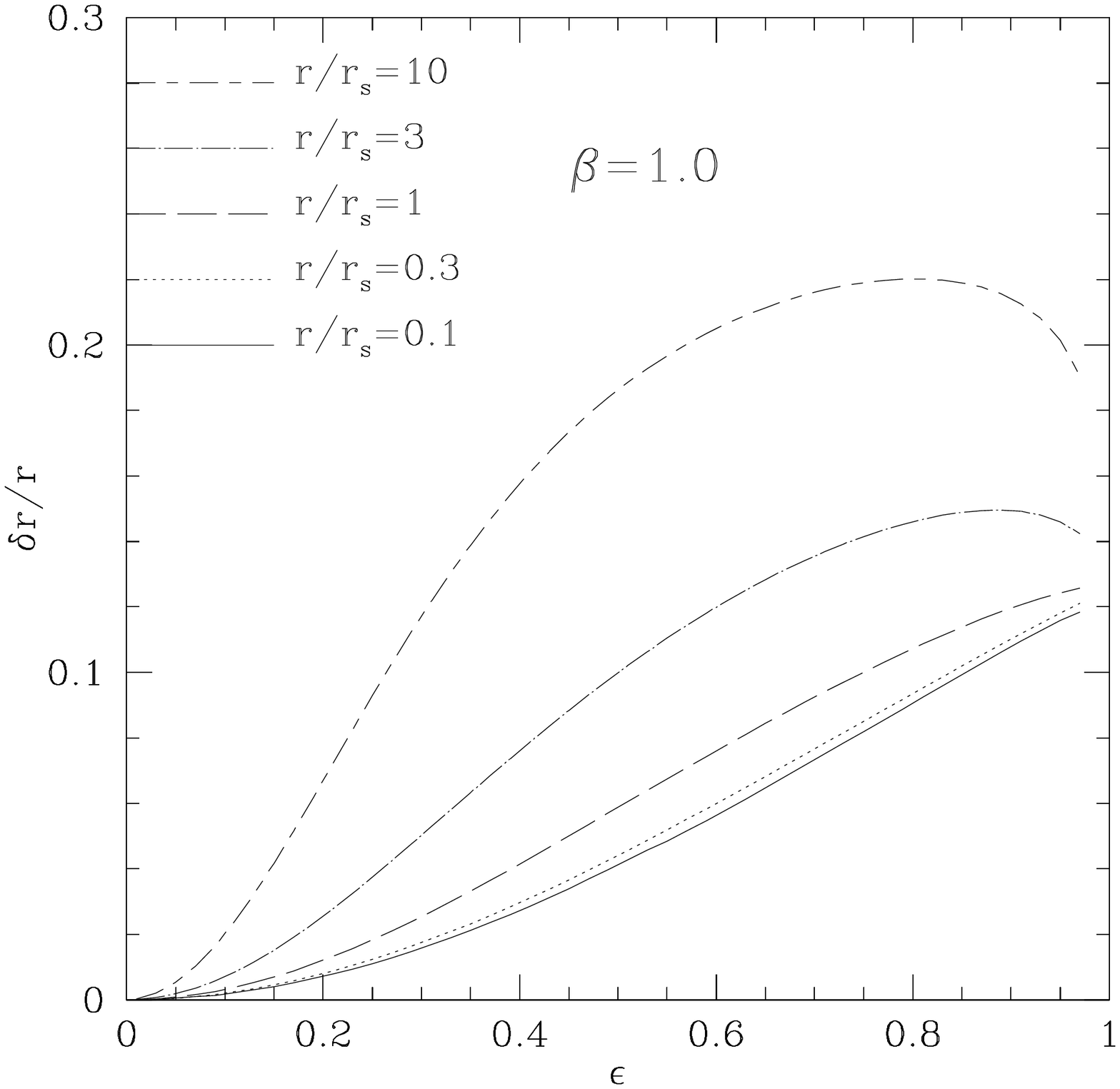}}
\mbox{\epsfysize=5.5cm \epsfbox{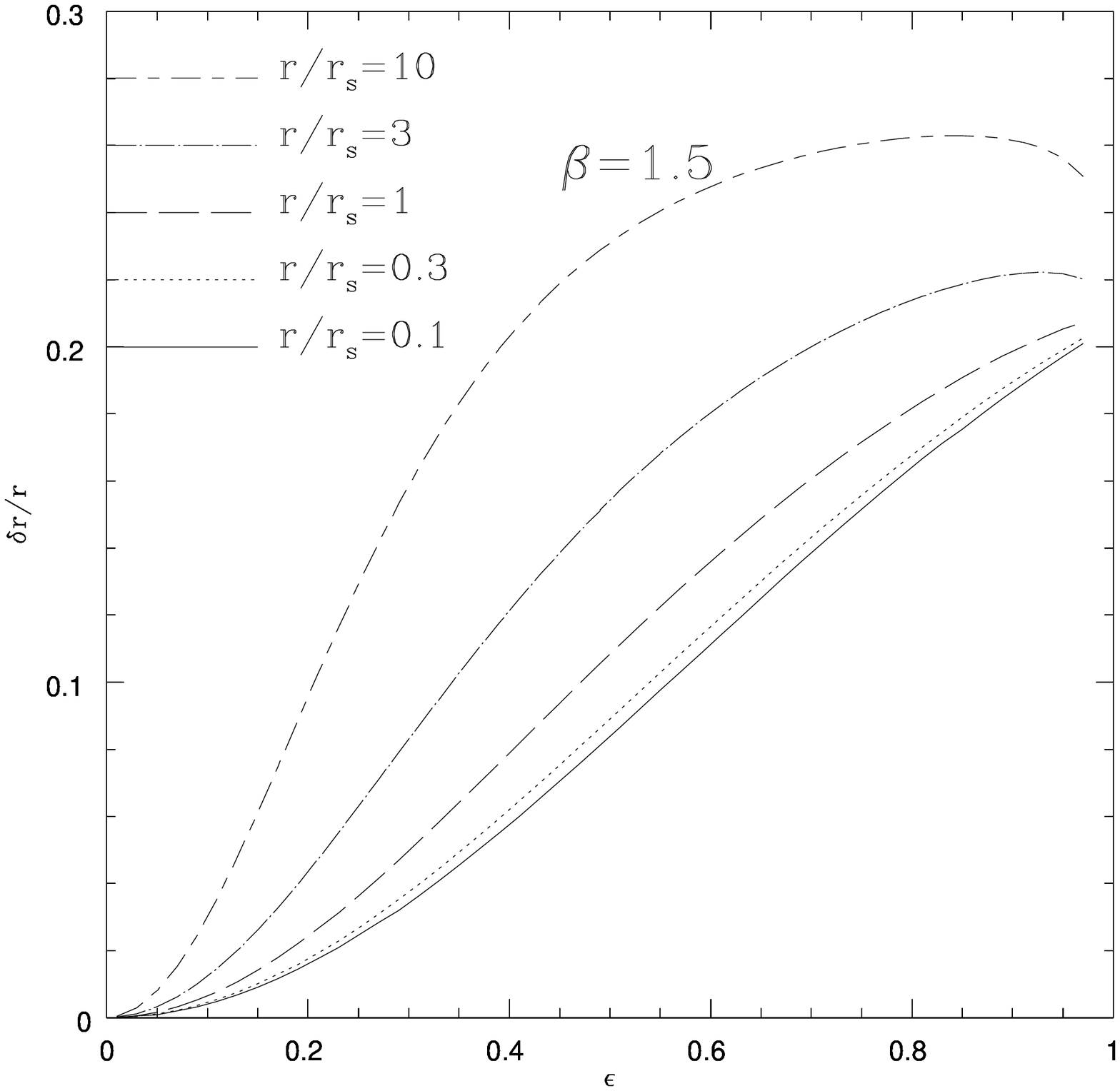}}
}
\caption{$\delta r / r$ as a function of $\epsilon$ for a variety of
pseudo-elliptical generalized NFW models with different inner slopes,
$\beta$.  This simply characterizes the deviation of the projected
density from an elliptical model for various $r / r_{sc}$.
\label{fig:delr}}
\end{center}
\end{figure*}

We quantify the degree of boxiness by measuring the distance,
$\delta r$, between a surface density contour and a real ellipse with
the same major and minor axis radii (as was done in \citet{Golse02};
see their Figure 6 for a geometrical illustration of $\delta r$).  In
Fig.~\ref{fig:delr} we plot $\delta r / r$ as a function of $\epsilon$
for several values of the inner slope, $\beta$, and a variety of
$r/r_{sc}$.  If we desire our pseudo-elliptical implementation to be
within $10 \%$ of a true elliptical surface density distribution for
$r/r_{sc} < 10$, then values of $\epsilon \lesssim 0.2$ are
appropriate, especially for DM halos with steep inner slopes.

\begin{figure*}
\begin{center}
\mbox{\epsfysize=6.0cm \epsfbox{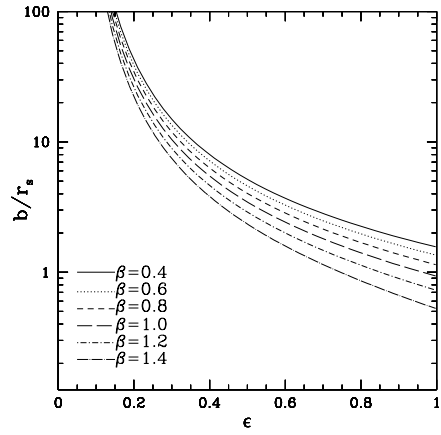}\label{fig:neg_test}}
\caption{Distance from ellipse center along the minor axis at which
$\Sigma_{\epsilon}$ becomes negative.  Several example values for
different inner dark matter density slopes, $\beta$, are plotted.}

\end{center}
\end{figure*}

One unphysical consequence of introducing ellipticity into the
potential is that the surface mass density can become negative,
especially near the minor axis where $cos(2 \phi_{\epsilon})=-1$.  In
Figure~\ref{fig:neg_test} we plot the distance along the minor axis at
which $\Sigma_{\epsilon}$ becomes negative for several inner slopes.
If we wish to restrict ourselves to values of $\epsilon$ where the
surface density does not go negative for $r/r_{sc} < 10$, then 
we must restrict ourselves to values of $\epsilon$ less than
approximately 0.25.

In summary, in order to be within $10 \% $ of a true elliptical surface
mass distribution and to have positive values of the surface mass
density for $r / r_{sc} < 10$ we must restrict our use of the
pseudo-elliptical gNFW parameterization to values of $\epsilon
\lesssim 0.2$, well within the model values of $\epsilon$ for the
clusters studied in this paper.

\bibliographystyle{apj}
\bibliography{apj-jour,mybib}

\end{document}